\documentclass[preprint2]{aastex}

\shorttitle{Disentangling the ICL with the CHEFs}
\shortauthors{Jim\'enez-Teja \& Dupke}

\begin{document}

\title{Disentangling the ICL with the CHEFs: Abell 2744 as a case study}
\author{Y. Jim\'enez-Teja\altaffilmark{1} and R. Dupke\altaffilmark{1,2,3,4}}
\affil{$^1$Observat\'orio Nacional, COAA, Rua General Jos\'e Cristino 77, 20921-400, Rio de Janeiro, Brazil}
\affil{$^2$Department of Astronomy, University of Michigan, 500 Church St., Ann Arbor,
MI 48109}
\affil{$^3$Department of Physics and Astronomy, University of Alabama, Box 870324,
Tuscaloosa, AL 35487, USA}
\affil{$^4$Eureka Scientific Inc., 2452 Delmer St. Suite 100, Oakland, CA 94602, USA}

\email{yojite@iaa.es}

\begin{abstract}
Measurements of the intracluster light (ICL) are still prone to methodological ambiguities and there are multiple techniques in the literature for that purpose, mostly based on the binding energy, the local density distribution, or the surface brightness. A common issue with these methods is the a priori assumption of a number of hypotheses on either the ICL morphology, its surface brightness level or some properties of the brightest cluster galaxy (BCG). The discrepancy on the results is high, and numerical simulations just bound the ICL fraction in present-day galaxy clusters to the range 10-50\%. We developed a new algorithm based on the Chebyshev-Fourier functions (CHEFs) to estimate the ICL fraction without relying on any a priori assumption on the physical or geometrical characteristics of the ICL. We are able to not only disentangle the ICL from the galatic luminosity but mark out the limits of the BCG from the ICL in a natural way. We test our tecnique with the recently released data of the cluster Abell 2744, observed by the Frontier Fields program. The complexity of this multiple merging cluster system and the formidable depth of these images make it a challenging test case to prove the efficiency of our algorithm. We found a final ICL fraction of 19.17$\pm$2.87\%, which is very consistent with numerical simulations.

\end{abstract}

\keywords{galaxies: clusters: intracluster medium, methods: data analysis, \objectname{Abell 2744}}

\section{Introduction} \label{intro}

 The diffuse intracluster light (ICL) is defined as the luminous component of the stars that are gravitationally bound to the cluster potential, but does not belong to any of the galaxies in the cluster. Thought to be mainly formed from the stripping of stars during the hierarchical process of accretion of the cluster \citep{willman,sommer-larsen,purcell,contini}, the ICL is the key to understand the formation history of the clusters as well as determining the correct baryon fraction for use in cosmology (e.g. \cite{allen,lima,lin}). The ICL colour, age, and metallicity  can provide valuable knowledge on the origin and evolution of the cluster, since they are directly connected with the characteristics of the progenitor galaxies \citep{trujillo}. Many studies have reported ICL colours as red as the elliptical galaxies in the cluster and that would suggest the ICL be primarily created at the early epochs of the cluster formation, during the build-up of the brightest cluster galaxy (BCG) and the other massive cluster galaxies \citep{krick, darocha, krick07}. However, other authors have found blue ICL colours which would point to accretion material stripped from satellite galaxies falling on to the cluster or from interacting cluster galaxies, thus indicating the ICL is mainly formed very late (z$<$1) \citep{zibetti,murante,contini,darocha,burke,trujillo}. In a few cases, intracluster star formation has been seen directly linking stripping to  ICL generation \citep{sun07,sun10}. Metallicity is also widely debated since some authors have reported metal-poor ICLs \citep{contini,durrell, williams,trujillo} whereas other works have found super-solar metallicities \citep{krick}. There is not an agreement on the spatial distribution of the ICL, since many studies have claimed it to be strongly aligned with the position angle of the BCG \citep{gonzalez,zibetti,presotto} while there are cases of remarkable misaligned tidal features, such as plumes, arcs, or tidal streams \citep{rudick09,krick,mihos,krick07}. \\
 
 As the ICL fraction (ratio between the ICL and the total luminosity of the cluster) is a much simpler parameter than the colour, metallicity or two-dimensional distribution, one would expect that an agreement on its estimates would be more easily achieved. Nevertheless, it is not understood how it relates to other physical properties of galaxy clusters. For instance, it is widely believed that this fraction correlates with the cluster mass \citep{zibetti,lin}, which would assume that ongoing processes as ram pressure stripping or harrasment are the dominant mechanisms of ICL formation. However, some studies have found constant ICL fractions as a function of the halo mass or a very weak dependence \citep{krick07,murante,contini}, which could be explained considering the galaxy-galaxy merging at the early stages of the cluster evolution as the primary mechanism. Theoretical works on ICL also show discrepancies on the final ICL fractions reported depending on the resolution of the numerical simulations \citep{murante,contini,cui}. Nonetheless, all authors agree that the ICL fraction likely increases with time, although the correlation would not be linear and it would depend on the dynamical evolution stage of the cluster \citep{krick,rudick,krick07,contini}.\\
 
 From the observational point of view, the separation of the ICL from the light of the stars locked up in the cluster galaxies is is very complex and so far there is not a standard method to unambiguously disentangle them \citep{rudick,contini}. Moreover, the definition of the transition from the BCG's extended outer profile to the ICL is not even tackled be many authors, who prefer to provide the ICL+BCG fraction \citep{pierini,presotto}. Methods employed in the literature are based on binding energy, local density or surface brightness. These techniques yield different results in the ICL fraction (up to a factor of ~4) using the same data \citep{rudick,cui}. The discordance extends also to the surface brightness methods and we can find different algorithms in the literature. The ICL morphology is sometimes a priori assumed by fitting a two-component profile (generally a double de Vaucouleurs profile or a double exponential profile or a combination of a de Vaucouleurs and a S\'ersic profiles) and identified with the most extended component \citep{gonzalez,zibetti,rudick}. Often, a surface brightness cut-off may be arbitrarily set, to disentangle the high-surface galactic luminosity from the low-surface ICL, so all the light over this threshold is masked out and the ICL fraction is measured from the remaining pixels \citep{krick07,zibetti,rudick,burke,cui,trujillo}. Other studies resort to masking all the galaxies in the cluster using a segmentation map or removing them using traditional profiles, assuming as ICL the remaining light \citep{mihos,williams,krick,presotto}. A more sophisticated approach is offered by \cite{darocha} or \cite{adami}, who perform a multiscale decomposition of the cluster elements, interpreting the low frequency component as the ICL. Therefore, it is not possible a direct comparison of the results, and this could explain the discrepancies observed in the literature.\\
 
 Our purpose is to develop an accurate, new method to dissociate the ICL from the galactic luminosity, including the BCG, not relying on any assumption on the ICL morphology, its surface brightness or the BCG profile. We use the CHEFs \citep{jimenez} and tools from differential geometry to implement our algorithm. Both are described in more detail below. As a case study we have chosen Abel 2744 cluster, not only for the superb data quality of the Frontier Fields program but also due to the challenging complexity this cluster offers. Abell 2744 is indeed a multiple cluster merger composed of at least four structures in essence being one of the most violent, active processes currently observed \citep{boschin,owers,merten,medezinski}. X-ray data support the existence of a post-core-passage major merger together with an interloping minor merger. With this scenario, an special technique to estimate the ICL fraction is mandatory, since it is needed to disentangle not only the galactic light from the ICL but also the intracluster brightness from the different subclusters in the system.\\
 
 This paper is organized as follows. Section \ref{data} describes both the imaging and spectroscopic data used, as well as the identification of the cluster member galaxies. Section \ref{chefs} outlines the mathematical background of the CHEFs which will be applied in several steps of the ICL algorithm, detailed in Section \ref{method}. We show the final results and draw the correspondent conclusions in Section \ref{conclusions}. Throughout the paper we assume a standard $\Lambda$CDM cosmology where $H_0$=70 km s$^{-1}$, $\Omega_m$=0.3, and $\Omega_{\Lambda}$=0.7. For the assumed cosmology and at the Abell 2744 redshift of z$\sim$0.3064, it yields 1''=4.53 kpc=4.15 h$^{-1}$ kpc.\\

\section{Data}\label{data}

 The cluster Abell 2744 (R.A.=00$^h$14$^m$18$^s_.$9, DEC=-30$^d$23$^m$22$^s$ [J2000.0], z$\sim$0.3064), also known as the Pandora cluster, appears to be the merging of four smaller galaxy clusters. The complexity and exceptional nature of the physical processes going on this system, make it a main target for optical, infrared, and X-ray imaging, as well as for spectroscopic campaigns. For this work we have used the very deep images from the main cluster of the Pandora system, just observed by the Hubble Space Telescope, along with the spectroscopy provided in \cite{owers}.
 
\subsection{Observations}

  Abell 2744 was observed by the Hubble Space Telescope (HST) within the frame of the Frontier Fields program ID 13495 (PI: J. Lotz). This program, started in 2013 to be ended by 2016, aims to obtain the deepest observations ever imaged of six selected clusters (Abell 2744, MACSJ0416.1-2403, MACSJ0717.5+3745, MACSJ1149.5+2223, Abell S1063 (RXCJ2248.7-4431), and Abell 370), as well as the second-deepest images from blank fields close to the main clusters. Data are imaged both by the Wide Field Camera 3 (WFC3) in the infrared bands Y, J, J+H, and H, and the Advance Camera for Surveys (ACS) with the optical filters B, V, and I. \\
 
 For the particular case of Abell 2744, the main cluster was observed in the infrared on October and November of 2013 with a total of 70 HST orbits, and in the optical from May to July of 2014 summing up 68 orbits. Observations did not include any of the three other surrounding clusters which compound the merging system of Pandora. These images were completed with the archival data from program ID 11689 (PI: R. Dupke), which provided other 16 orbits of observation in the optical filters, along with another orbit in the infrared from program ID 13386 (PI: S. Rodney). The individual exposures were calibrated using the standard STScI pipeline, which corrects for bias and dark current substraction, flatfield correction, electronic gain calibration, non-linearity and CTE corrections, bad pixel rejection, and photometric calibration. The images are visually inspected and recalibrated if necessary, to be later aligned to compose the final mosaics\footnote{Further information on HST reduction pipeline can be found in http://archive.stsci.edu/pub/hlsp/frontier/abell2744/images/ hst/v1.0-epoch2/hlsp\_frontier\_hst\_abell2744\_v1.0-epoch2\_readme.pdf}. The final dataset has depths at 5$\sigma$ in a 0.4" diameter aperture ranging between $\sim$27.1-28.7 magnitudes [AB] accross all filters \citep{laporte}. The surface brightness limits at 3$\sigma$ range between $\sim$29.13-30.32 mag/arcsec$^2$ \citep{trujillo}.\\

 In this work we have used the F814W band from Abell 2744 at 60 mas/pixel, retrieved from the Frontier Field archive\footnote{http://archive.stsci.edu/pub/hlsp/frontier/abell2744/images/hst/} (see Fig. \ref{orig&bg}(top)). We have chosen this band because the ICL in this cluster has a bluer colour than expected \citep{trujillo}. The image has a total area of 11.9 arcmin$\mathbf{^2}$ ($\sim$930x940 kpc), summing up a total of 46 HST orbits. \\
 
\subsection{Spectroscopy and cluster membership}

 \cite{owers} produced a catalog with redshift estimations of 1443 sources in the Abell 2744 field using the AAOmega MOS on the Anglo-Australian Telescope, combined with the existent data in the literature.  The quality of the estimated redshifts is indicated in the combined catalog, so we have been able to select a total of 1237 extragalactic objects with highly robust redshifts within a 15 arcmin radius of the center of the main cluster.\\
 
 We have complemented the catalog by \cite{owers} with the spectroscopic redshifts available in the Nasa Extragalactic Database\footnote{https://ned.ipac.caltech.edu/} (NED). Our combined catalog contains precise redshift estimations of 1518 sources in the Pandora system, which we have used to determine the cluster membership of the galaxies.\\
 
\begin{figure}[h!]
\centering
\includegraphics[width=7.8cm]{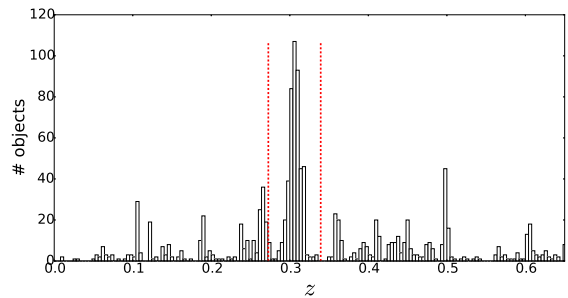}
\caption{Redshift distribution of the objects in the combined catalog. Dashed vertical lines encircle the redshift window of possible cluster members, as identified by the PEAK method.}\label{peak}
\end{figure}

 We have used a two-step analysis to identify the cluster members, based on the PEAK and the shifting gapper methods \citep{fadda,owers}. The PEAK method \citep{fadda, girardi,boschin} roughly discards interloping galaxies and groups studying their distribution in redshift. It consists on identifying the peak of the redshift histogram (usually matching the mean cluster redshift, $z_{cluster}$) to later point out the galaxies within a window around this peak as possible cluster members. Similarly to \cite{owers} and given that this cluster has evidence for merging, which could enhance the velocity range, we have chosen the window to be $c\,z_{cluster}\pm 10000$ km/s. In Fig. \ref{peak} we can observe the redshift distribution of the objects in our combined catalog in redshift intervals of 0.004. The histogram peaks at $z\sim$0.304, what was expected since the mean cluster redshift is estimated to be $z\sim$0.3064. The window allocating the possible cluster members is enclosed by the dotted vertical lines in Fig. \ref{peak}.\\
 
 The shifting gapper method \citep{fadda, owers, boschin} is used to refine the crude selection made by the previous technique. It is based on the fixed-gap method \citep{zabludoff, fadda, biviano} which analyses the line-of-sight velocity-projected clustercentric distance space. The combination of both redshift and spatial information permits a better determination of the interlopers and decreases the contamination of the selection. The shiffting gapper method firstly splits the galaxy sample in radial bins from the center of the cluster. Within each spatial bin, the objects are sorted according to their peculiar velocity, defined as:
 
\begin{equation}
v_{pec}=c\,\frac{z-z_{cluster}}{1+z_{cluster}}.
\end{equation}

 We have considered all those objects with positive peculiar velocity $v_{pec}$, sorted them, and calculated the differences in peculiar velocity for every two contiguous objects (the so called velocity gaps). The first objects with an associated velocity gap greater than the velocity dispersion of the sample are rejected, as well as all the subsequent objects in the sorted list. The same process is applied to the objects with negative peculiar velocities. This analysis is iterated for each spatial bin. To compute the statistical parameters, we have used the biweight estimators described in \cite{beers}, since they have been proved to perform better, especially in the case of non-Gaussian or contaminated normal distributions.\\
 
\begin{figure}[h!]
\centering
\includegraphics[width=7.8cm]{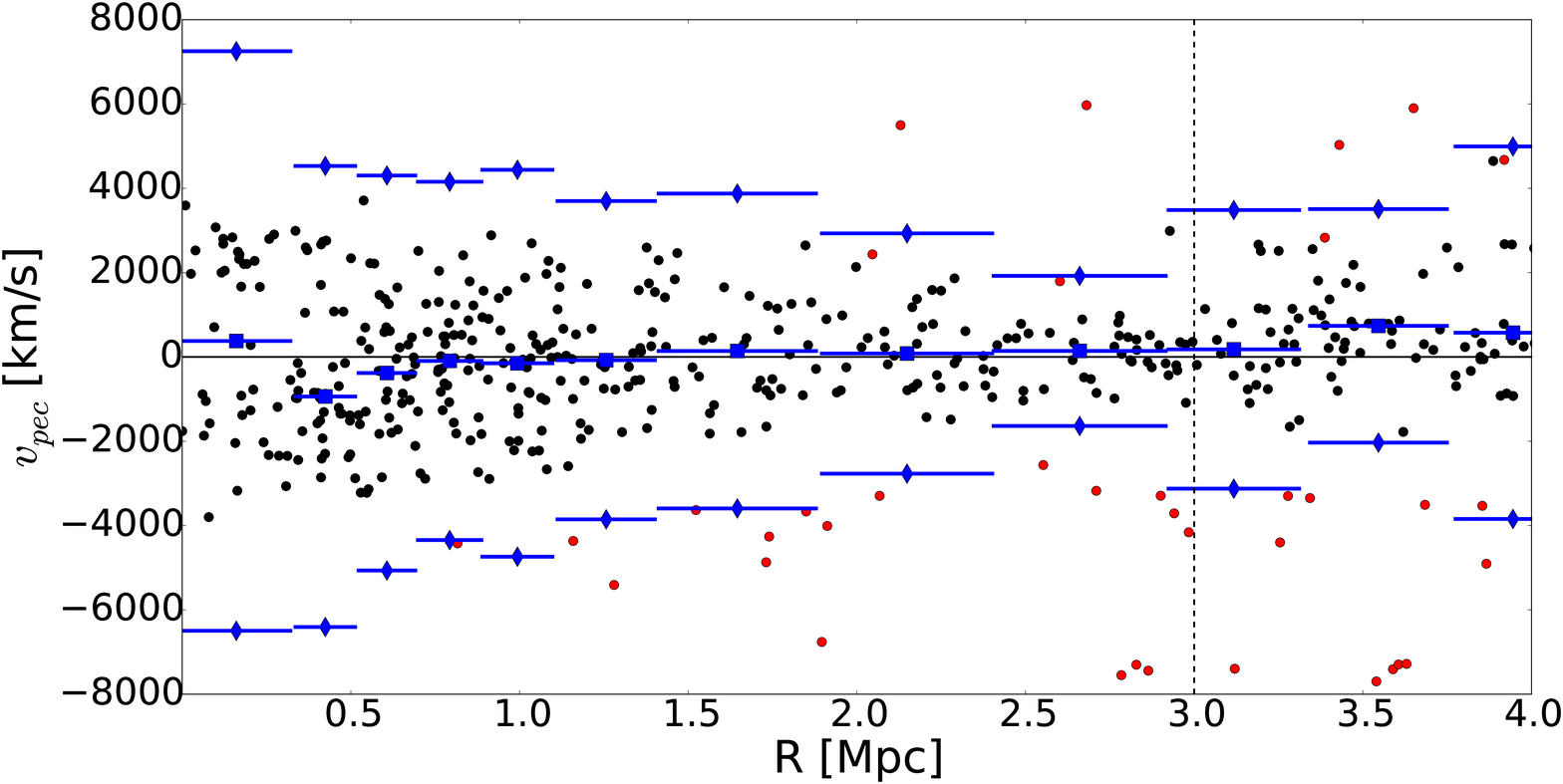}
\caption{Shifting gapper diagram. We plot the objects remaining after the crude separation by the PEAK method. Black dots correspond to galaxies selected as cluster members while the red ones are rejected. Blue horizontal lines with squared markers indicates the mean of the distributions in the radial bins, while the blue lines with diamond markers show the 3-$\sigma$ limits.}\label{shifting_gapper}
\end{figure}

 We show the peculiar velocity versus projected clustercentric distance diagram in Fig. \ref{shifting_gapper}. As in \cite{owers}, the shifting gapper method is only valid for the galaxies within a 3 Mpc radius from the center of the cluster, since beyond that distance the separation is not clear. Our analysis yielded a final number of 348 cluster members within this radius (versus the 343 found by \cite{owers}), with 78 of them lying in the 11.9 arcmin$\mathbf{^2}$ field imaged by the Frontier Fields.\\
 
\section{CHEFs mathematical background}\label{chefs}

 The Chebyshev-Fourier functions \citep[CHEFs, ][]{jimenez} are a mathematical tool that was originally designed to model the light surface distribution of the galaxies. Each CHEF function is composed by a Chebyshev rational function $TL_n(r)$ of order $n$ and a Fourier mode $W_{m}(\theta)$ with frequency $m$ (this notation stands for both $\sin{(m\theta)}$ and $\cos{(m\theta)}$), in polar coordinates $(r,\theta)$:
 
\begin{equation}\displaystyle
\left\{\phi_{nm}(r,\theta;L)\right\}_{nm} =\left\{\frac{C}{\pi}\;TL_{n}(r,L) 
W_{m}(\theta)\right\}.\label{basis}
\end{equation}

 where $C$ is just a normalization factor, with value $C=1$ for $n=1$ and $C=2$ otherwise. This set of functions, depending on the so called scale parameter $L$, constitutes a basis of the space of the bidimensional, smooth, square-integrable functions. That means that we have a different basis for each value of $L$, able to fit the two-dimensional light distribution of any galaxy in the sky. The effectiveness of a basis, that is, the number of CHEF elements required to efficiently model an object within a certain level of accuracy, depends on the optimization of the scale $L$. This parameter expresses the level of compression or dilation of the CHEF functions in the plane, thus it is related to the size of the object being fitted (see \cite{jimenez2} for further details). Once the value of $L$ is set, the decomposition of the light surface distribution of a galaxy $f(r,\theta)$ into a linear combination of CHEFs is:
 
\begin{equation}\displaystyle
f(r,\theta)=\frac{C}{2\pi^2}\sum\limits_{m=0}^{\infty}\sum\limits_{n=0}^{\infty} f_{nm}\, TL_n(r)W_m(\theta),
\end{equation}
 
 where $f_{nm}$ are the so called CHEF coefficients. As the CHEF bases are orthonormal, these coefficients are calculated by means of an inner product:
 
\begin{equation}
\displaystyle f_{nm}=\frac{C}{2\pi^2}\int\limits_{-\pi}^\pi
\int\limits_0^{+\infty} f\left(r',\theta'\right)TL_{n}\left(r'\right)W_m\left(\theta'\right)
\frac{1}{r'+L}\sqrt{\frac{L}{r'}}\;dr'\;d\theta'.\label{coeffs}
\end{equation}

\section{ICL disentangling with the CHEFs}\label{method}

 The accuracy in the CHEFs models makes them ideal to be applied to many different study fields apart from the morphology, as for instance, photometric measurements \citep{jimenez2}. In this work we will use the CHEFs not only to fit the galaxies in the cluster images, but also to disentangle the luminous bulk of the BCG from the ICL. We have used the new, extremely deep image from cluster Abell 2744 to test our technique. The analysis consisted of six basic steps: removing all the sources in the image using the CHEFs including the BCG, re-adding the CHEF model of the BCG, determining the real extension and shape of the BCG by delimiting the points where the curvature of the surface BCG+ICL changes (i.e., the points where the ``slope'' of the surface changes), refitting the BCG with the CHEFs constraining the model to the region previously determined, estimating the background of the image, and measuring the resulting ICL.\\
 
\subsection{Removing the objects with CHEFs} \label{remo}

\begin{figure}[h!]
\centering
\includegraphics[width=7cm]{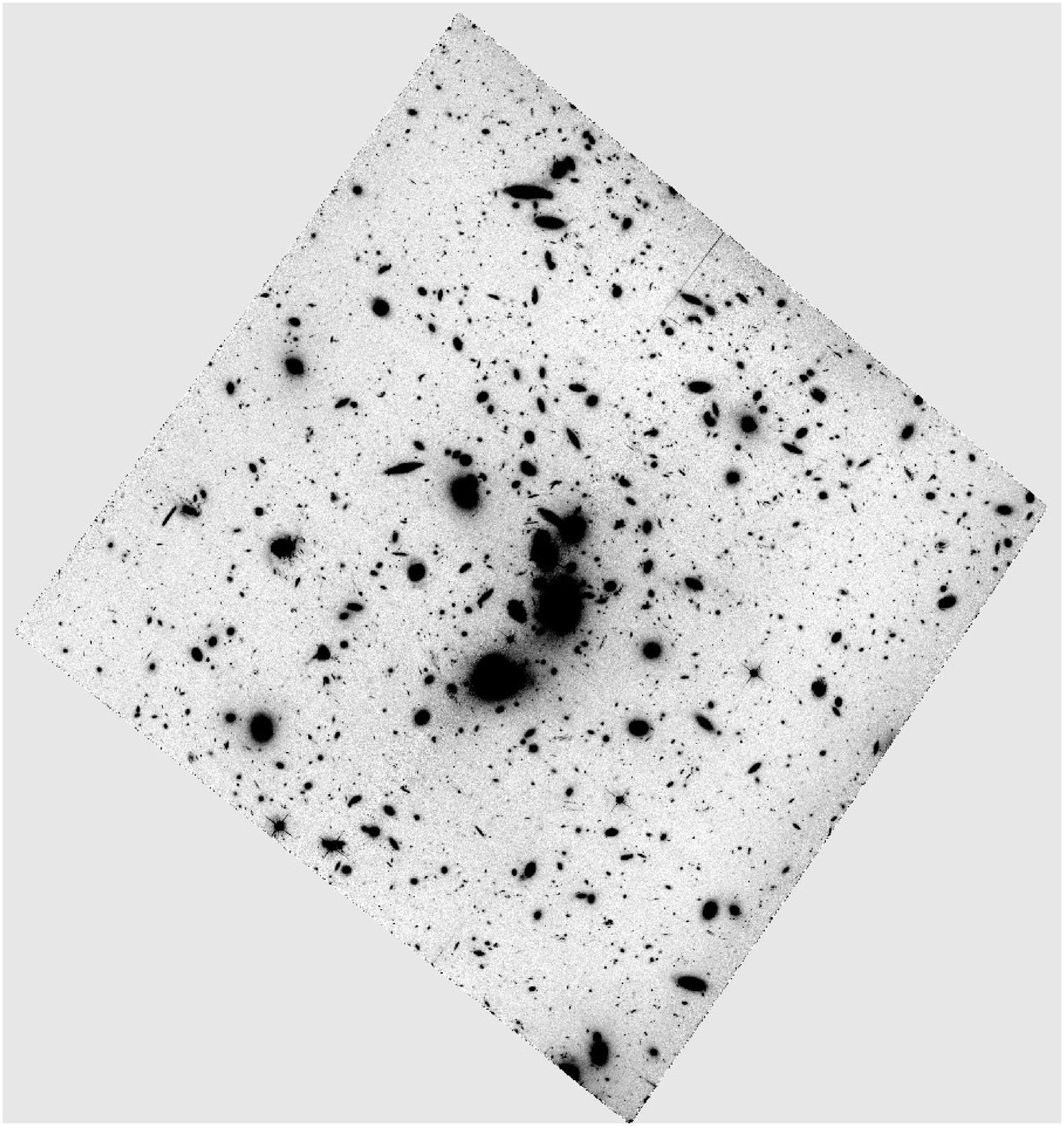}
\includegraphics[width=7cm]{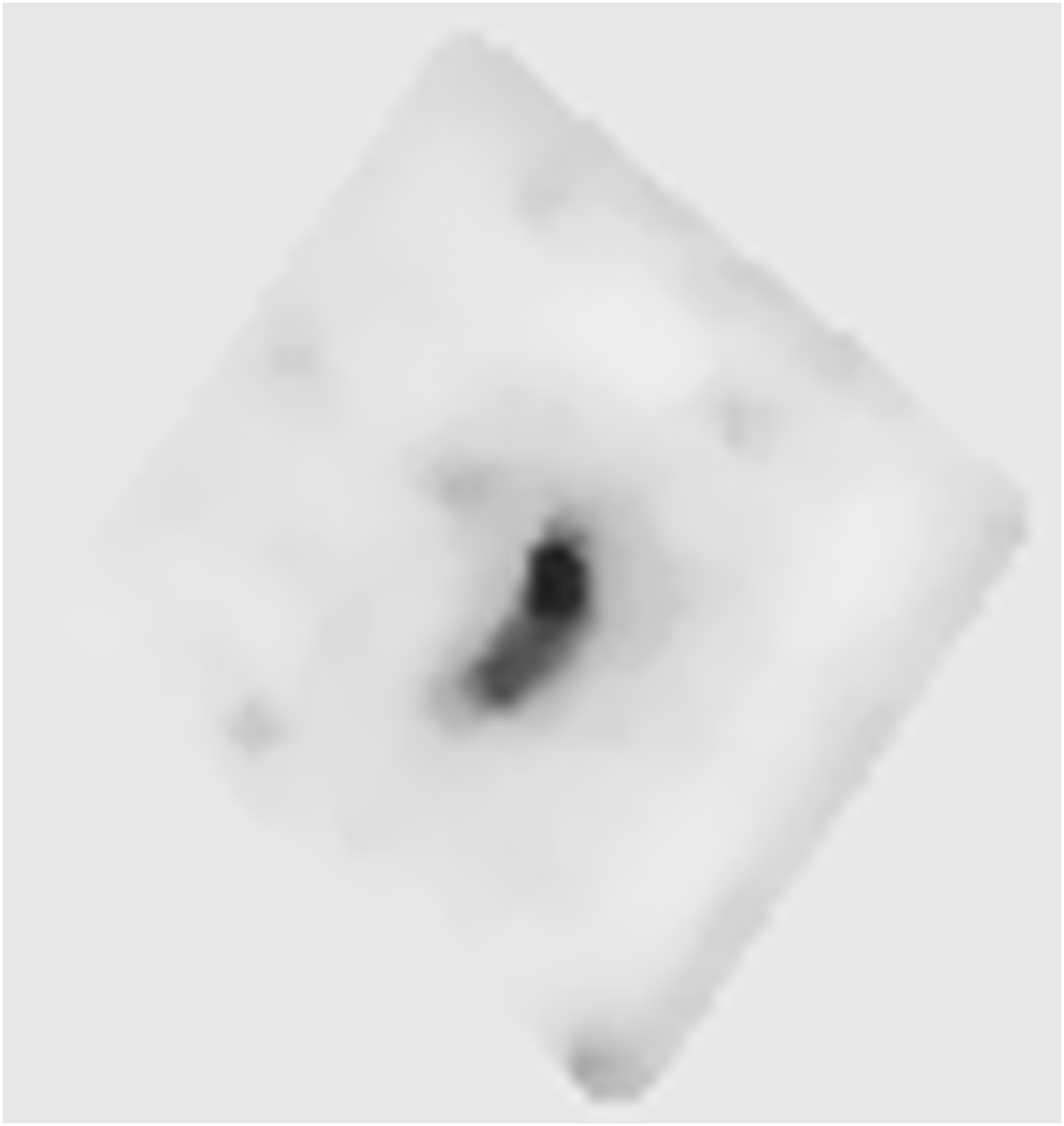}
\caption{ Original F814W band from Abell 2744 as observed by the Frontier Fields program (top) and SExtraction estimation of its background (bottom) with the configuration described in Sect.\ref{remo}.}\label{orig&bg}
\end{figure}

 As the CHEF functions constitute mathematical bases, by definition they are able to fit any galaxy. Just saturated stars lie out of CHEFs' reach, since they are not smooth and thus do not belong to the mathematical space of functions modeled by the CHEFs. So the first steps to process any image with the CHEFs is finding an optimal SExtractor \citep{sextractor,sextractor2} configuration to properly detect the sources and masking all the stars. CHEFs can fit unsaturated stars but we preferred to mask them out too to prevent the algorithm from trying to model any saturated star unnecessarily wasting computing time; CHEF models of saturated stars are very large due to their long diffraction spikes, which makes them computationally very expensive. For the particular case of Abell 2744, we have chosen to run SExtractor with a detection threshold of 2$\sigma_{sky}$ and a minimum area of 3 pixels. The background level and rms were determined locally, setting BACK\_SIZE=64, BACK\_FILTERSIZE=5, and BACKPHOTO\_THICK=24 in SExtractor parameters. The background map produced by Sextractor is a bi-cubic-spline interpolation evaluated on a mesh of size BACK\_SIZE. If this parameter is small, part of the flux of the larger objects can be absorbed in the background. We purposely chose a BACK\_SIZE greater than the average size of the galaxies but much smaller than the extended ICL, to make SExtractor consider this ICL as part of the background (see figure \ref{orig&bg}). Our goal was ensuring we were not going to remove any light from the ICL component in the subsequent steps of the algorithm. In this way, the CHEFs would just model the flux above this background level, remaining the ICL intact to be measured later. With these settings, the BCG light was also partly included in the background estimation because it was larger than 64x64 pixels. To remove this contamination we processed the BCG separately in the following steps of the analysis. \\
 
\begin{figure*}
\vspace*{-2cm}
\centering
\epsscale{2.}
\setbox1=\hbox{\plottwo{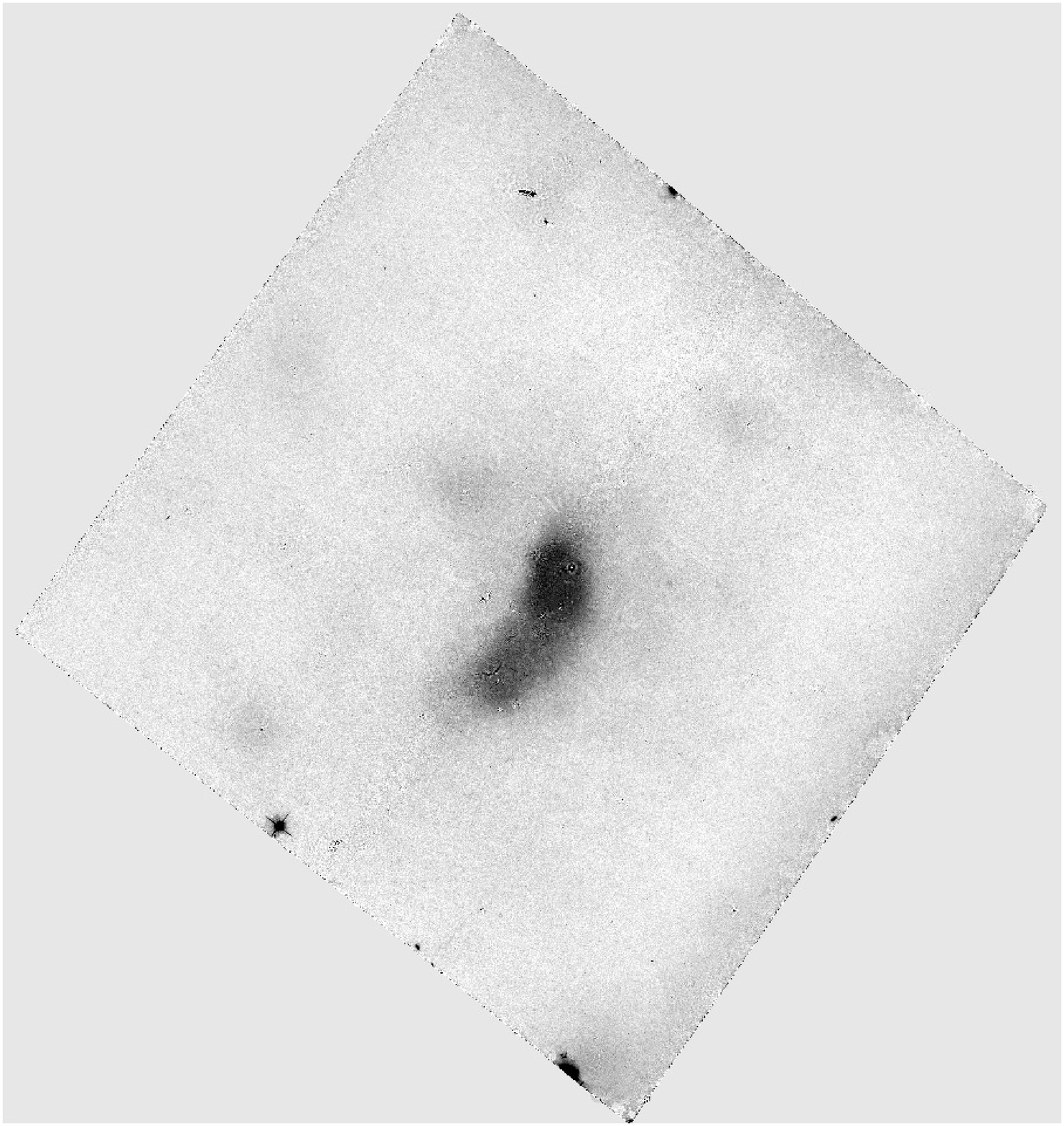}{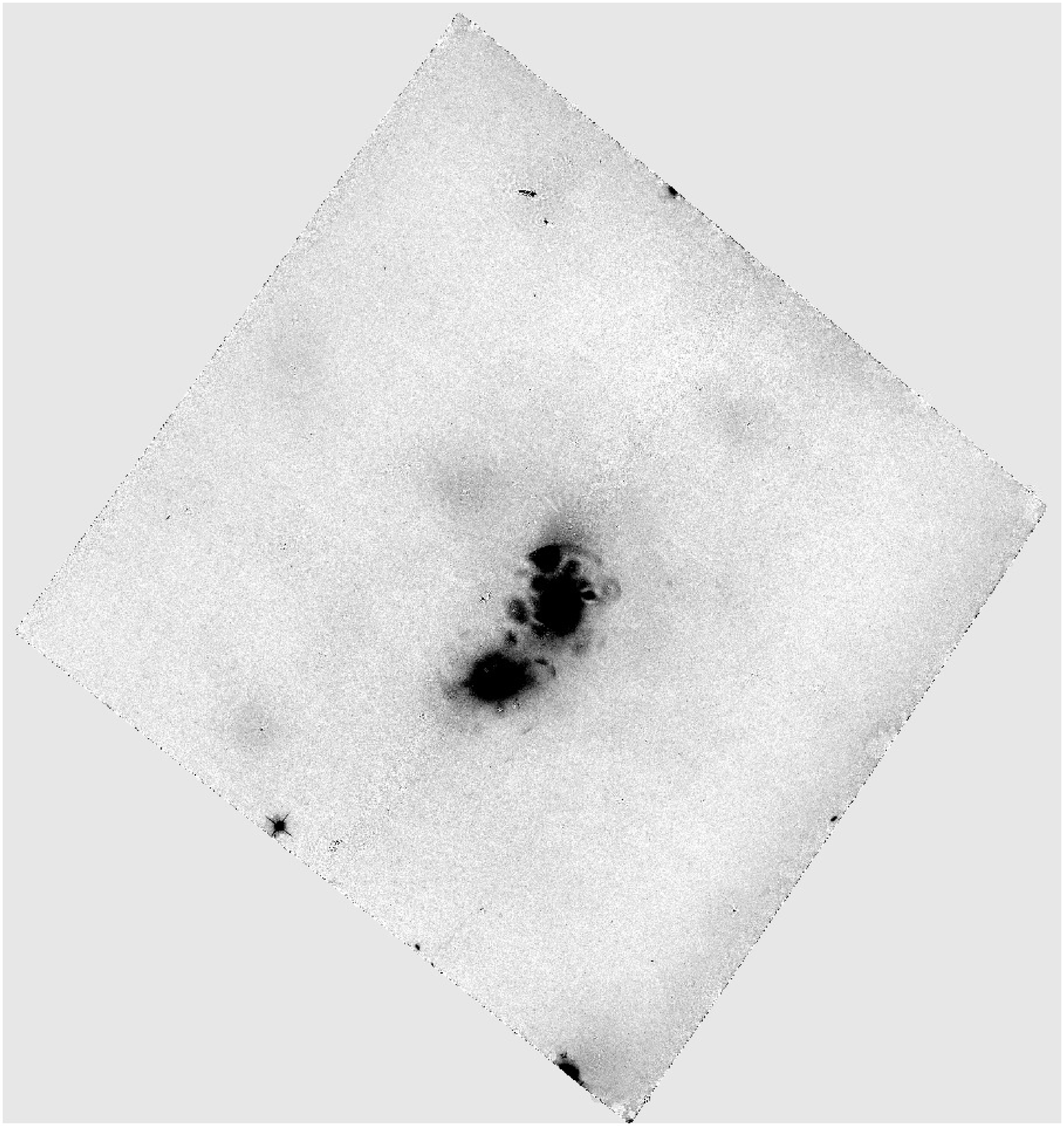}}
\plottwo{2residual.eps}{5Residual+BCGs.eps}
\llap{\makebox[\wd1][l]{\hspace{-0.7cm}\raisebox{0.3cm}{\Large{(a)}}}}\llap{\raisebox{0.2cm}{\Large{(b) }}}\\
\plottwo{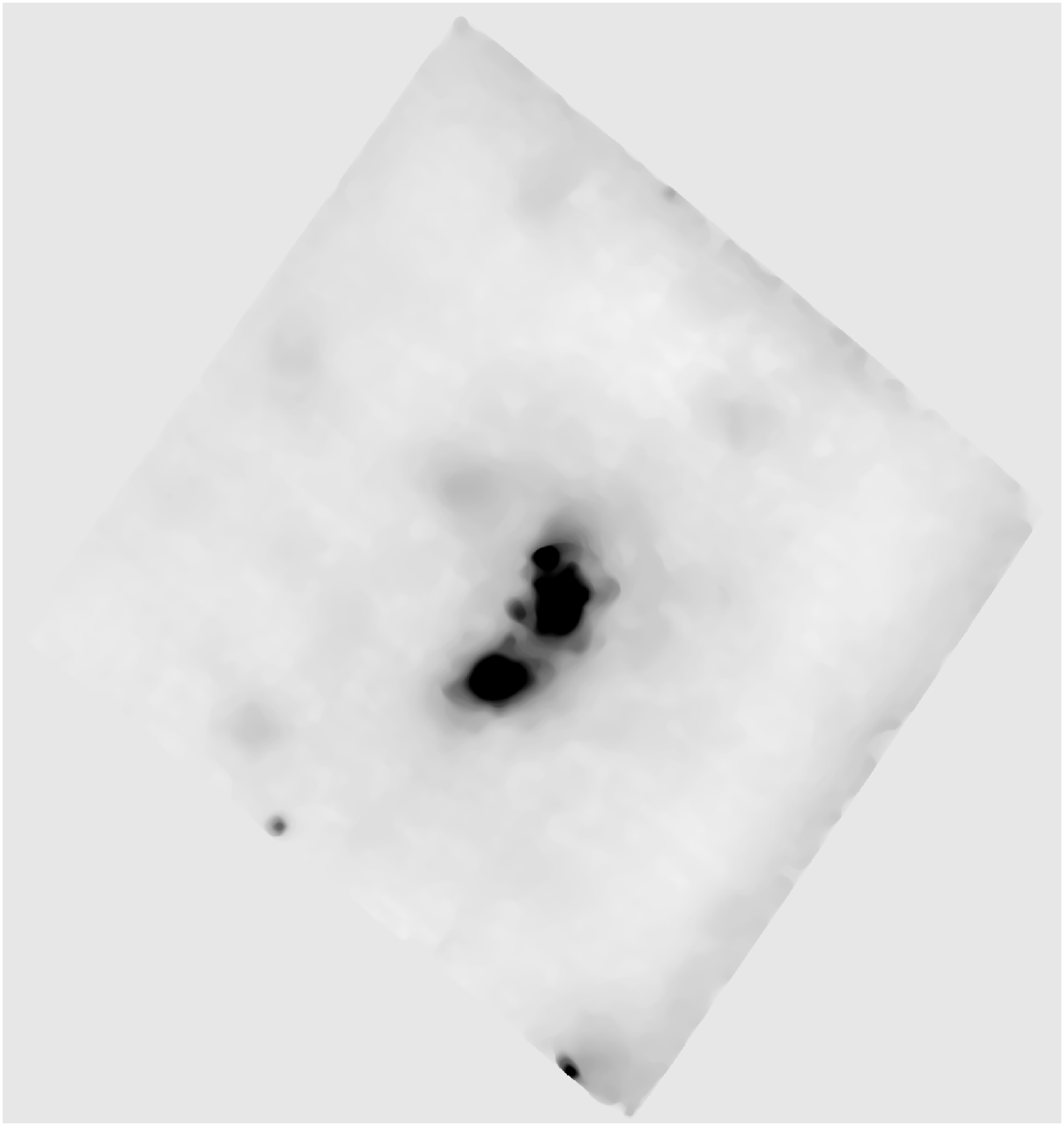}{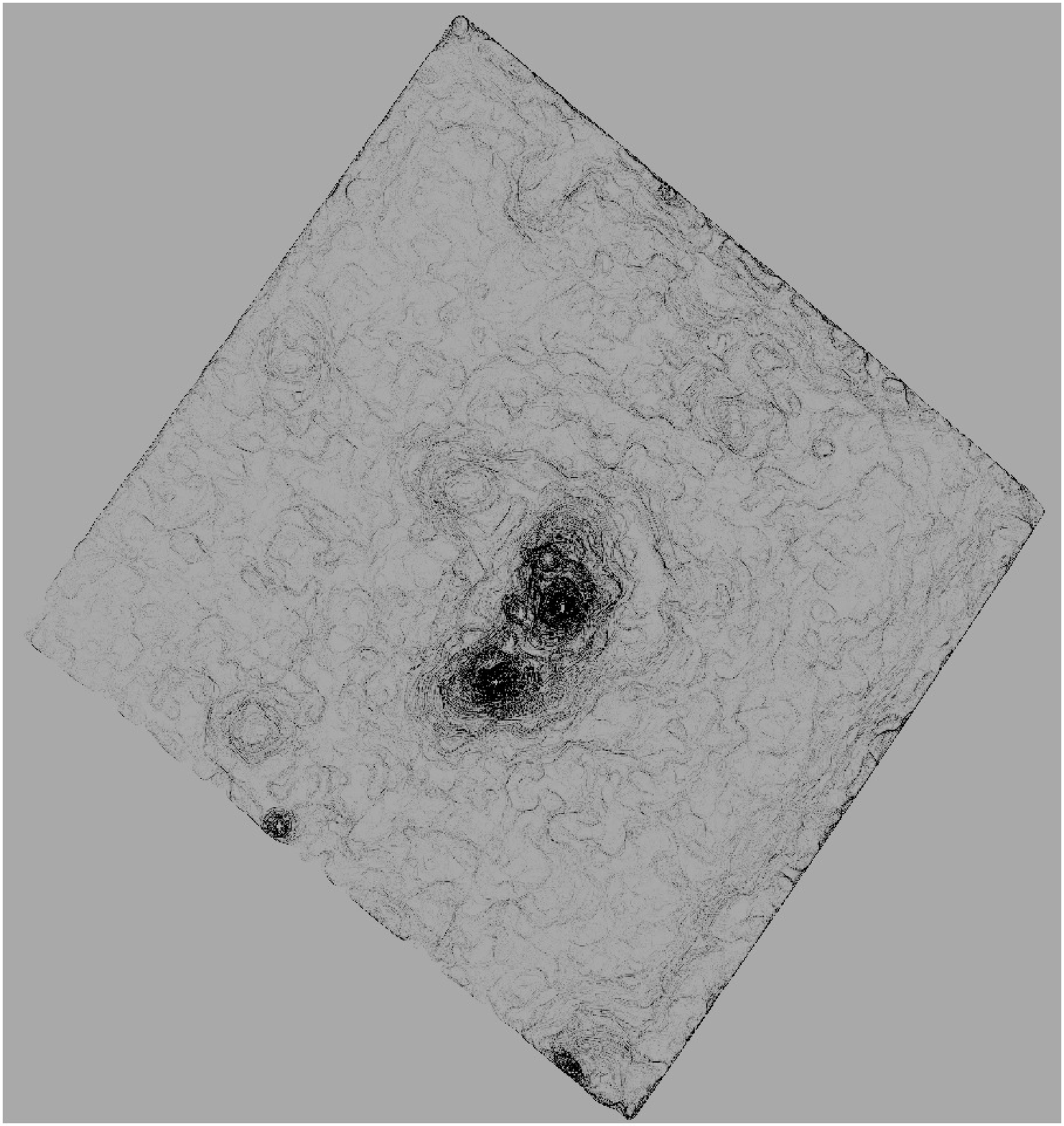}
\llap{\makebox[\wd1][l]{\hspace{-0.7cm}\raisebox{0.3cm}{\Large{(c)}}}}\llap{\raisebox{0.2cm}{\Large{(d) }}}\\
\plottwo{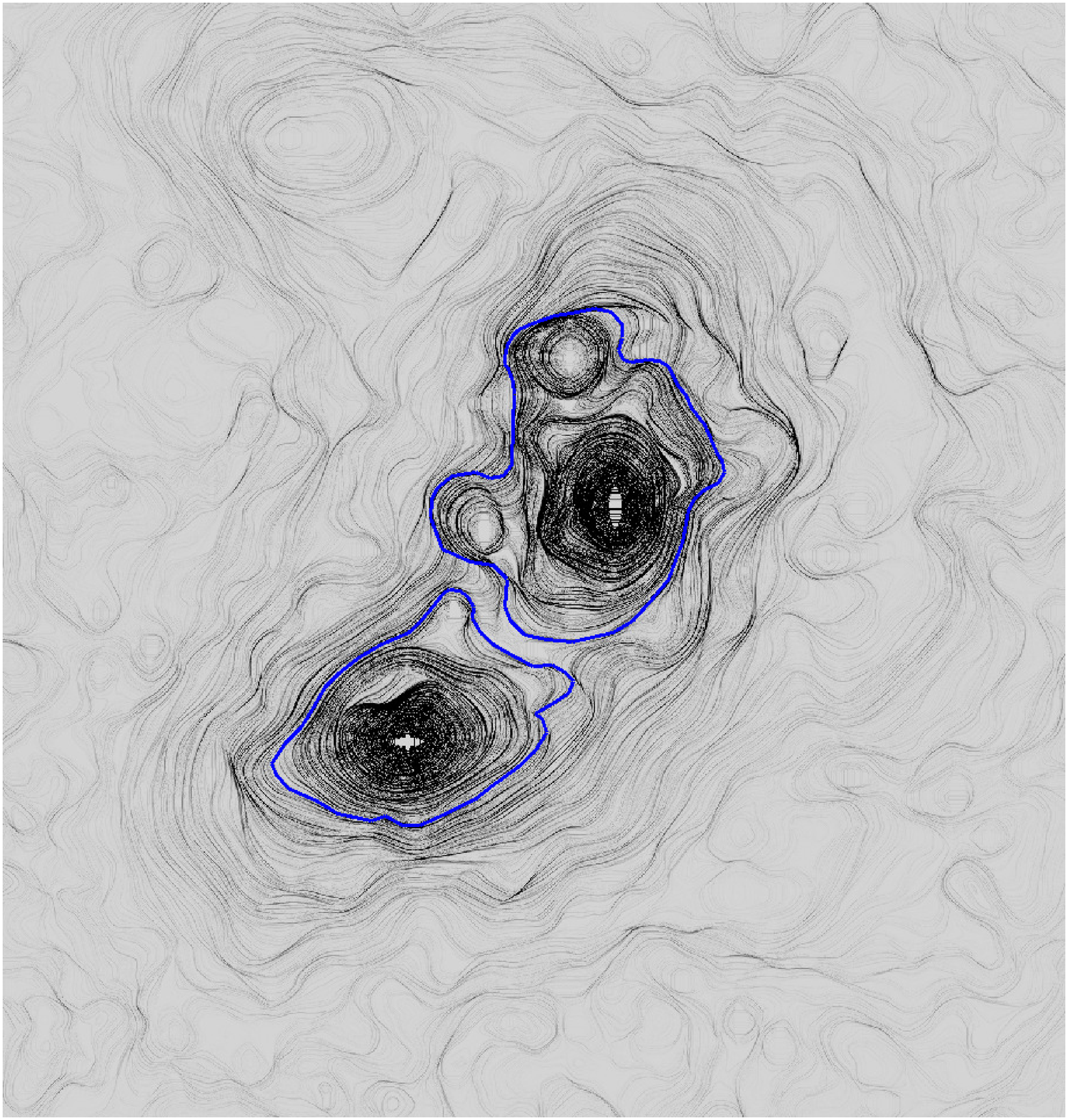}{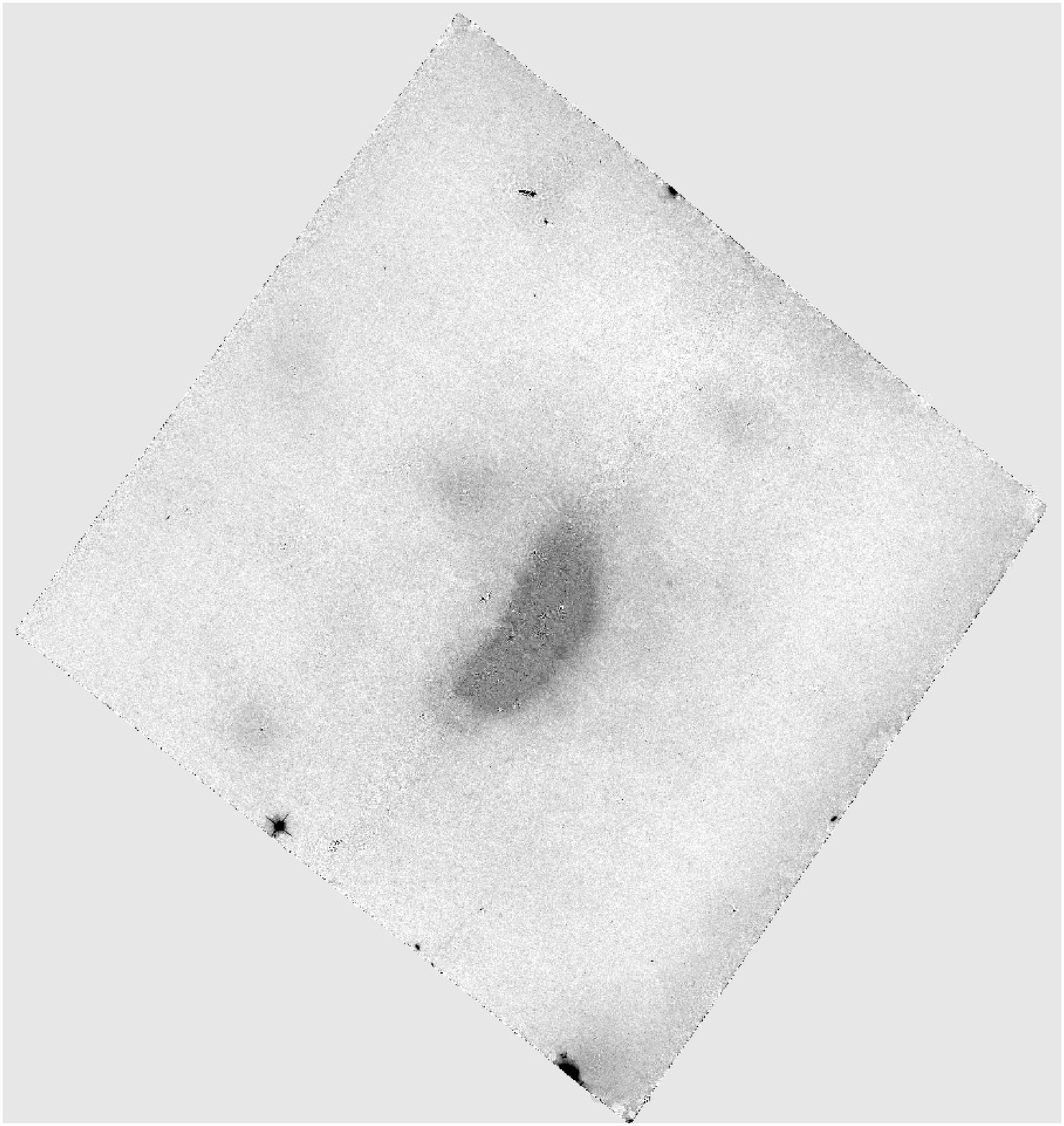}
\llap{\makebox[\wd1][l]{\hspace{-0.7cm}\raisebox{0.3cm}{\Large{(e)}}}}\llap{\raisebox{0.2cm}{\Large{(f) }}}
\caption{Analyses of Abell 2744 with the CHEFs. We show the different stages in the process of disentangling the ICL: a) residual image after removing all the objects with the CHEFs (remaining objects are stars or galaxies which did not achieve a good enough $\chi^2$ in the fitting algorithm); b) previous residual image with the brightest galaxies in the central area re-added (this region thus remains as in the top panel of  Figure \ref{orig&bg}); c) denoised version of (b) using the PMT at scale 6; d) MPC map of the clean BCG+ICL+background surface in (c); e) close up of the central region in (d) with limits marking out the BCGs extension; and f) final residual image after subtracting the new CHEF models of the BCGs, constrained to the regions shown in (e). This last image is just composed of ICL and background.}\label{CHEFalg}
\end{figure*}

 The CHEFs were run using the SExtractor background and rms background maps estimated as described.  A maximum number of CHEF coefficients of $n=m=15$ was allowed, except for the case of the two largest galaxies of the image, for which we permit an upper limit of $n=m=20$. Please note these limits do not express the final number of coefficients used for the CHEF models, whose optimal values are internally calculated by the CHEF algorithm. These numbers indicated the maximum values permitted. They exclusively depend on the size and substructure content of the galaxies, and this setting is found to reach a good trade-off between computing time and the accuracy of the resulting models (see \cite{jimenez} for further details). With this configuration we have obtained the residual image shown in Fig. \ref{CHEFalg}(a). As can be seen, there is still much light in the central area and at this level of reduction we are not able to discern whether it belongs to the BCG, the ICL or the background. As we mentioned above, this fact is not important at this stage since we will later re-add the CHEF BCG model to the image to make a closer inspection and analysis of this object and efficiently separate its contribution from the light surface distribution in that central area.\\
 
\subsection{Re-adding the BCG model}

 All the output, individual CHEF models were stored and the information concerning them (scale parameter $L$, centroid, size of the frame in which the CHEF model was computed, and number of Chebyshev $n$ and Fourier $m$ coefficients used) was saved in a catalogue. So it is straightforward to reinsert the CHEF model at its corresponding location in the original image. With this process we completely recovered the central area of the image, as in the original data. For the particular case of Abell 2744, we noticed several bright, extended galaxies in the central area of the cluster, so we decided to treat them all as BCGs. In Fig. \ref{CHEFalg}(b) we show the original image with all the objects removed except for the BCGs in the central area of the cluster. This image includes background, ICL, and the BCGs luminosity.\\

\subsection{Delimiting the BCGs} \label{delimiting}

 The CHEFs, as mathematical bases, attempt to model everything considered signal down to the background noise. The ICL would thus be fitted by the CHEFs as part of the BCGs. To avoid that, we initially marked out the BCGs ``real'' extension to later constrain the CHEFs to that area.  This is the real challenge in ICL measurement, since it is very complex to disentangle the BCG light from the ICL. \\
 
 We use the concept of curvature to analyze the BCG+ICL surface and separate the two components. The curvature is a bidimensional map of the intensity steepness of a surface. It is thus a very powerful tool to disentangle different components in a surface, just assuming each component has a different slope. \\
 
 Mathematically, the curvature of a curve quantifies the degree of dissimilarity of the curve from a straight line. It is calculated by comparing the different slopes of the tangent lines of nearby points in the curve. Therefore, its value is an intrinsic characteristic of each point on the curve. The curvature of a surface generalizes this concept to two dimensions gathering, for each point on the surface, the curvature of the infinite set of lines embedded in the surface and passing through that point. Intuitively, the curvature of a surface expresses the "slope" of the surface on each point.\\
 
 We assume the BCG+ICL surface light distribution suffers a change in the curvature at the points delimiting the galaxy flux, since we expect the BCG surface brightness to have a different ``slope'' than the ICL. This is the only assumption of the method and it would be expected to be the case for the ``true'' diffuse ICL, and could be the case even for the diffuse intrahalo stars (IHL). Although other physical effects could mimic a change in the curvature (such as high star formation areas, spiral arms or tidal streams, for instance), these effects would be local and easily distinguishable from the global change in curvature produced by the ICL. In other words, since the ICL is associated to the low frequency light distribution components, we would expect it to have a much smaller (or at least different) ``slope" than that of the BCG light distribution. In fact, the idea is quite similar to that assumed by the works where a two-component profile is fitted to the BCG+ICL surface \citep{gonzalez,zibetti,rudick}, with the advantage of not assuming any a priori hypothesis on the shape of the profiles.\\
 
 The idea is then to calculate a curvature map of the BCG+ICL surface and identify the points where this curvature changes. Among all the existent curvature parameters in differential geometry, we have chosen the Minimum Principal Curvature (MPC, \cite{diff_geom}), which is defined as follows. Given a point $P$ on a surface, we calculate the curvatures of the whole set of curves embedded in this surface and passing through $P$. It can be proved that these values range from a minimum value $k_1$ and a maximum $k_2$, which are the so called Minimum and Maximum Principal Curvatures, respectively. Analytically, these two parameters can be derived as the solutions of the quadratic equation:
 
\begin{equation}\displaystyle
(EG-F^2)k^2+(2FM-EN-GL)k+(LN-M^2)=0\label{curvatura}
\end{equation}

 where $E$, $F$, and $G$ are the components of the First Fundamental Form, and $L$, $M$, and $N$ the components of the Second Fundamental Form \citep{diff_geom}. Among the various possible curvature parameters (Gaussian curvature, Mean curvature, etc), the MPC seemed to be the most sensitive to changes in the ``slope'' of convex surfaces. \\
 
 However, it must be noted that the MPC is a geometrical characteristic of each point on a surface, so we will have a map of curvatures for the complete image. As our BCGs+ICL+background image contained noise, calculating the MPC map directly from these data produced a non-smooth map where it was not possible to discern the points where the curvature changed. We have chosen the Pyramidal Median Transform (PMT, \cite{starck}) to  filter the BCGs+ICL+background image and decrease the noise before computing the MPC map. This transform was specially designed by \cite{starck} to compress and denoise astronomical images, and it allows to compress data without loss of information in different levels that can be easily identified as noise and signal. Mathematically, the PMT is a multiscale transform which convolve an image iteratively with a kernel performing the median in a box of a certain size. The larger the number of iterations, the smoother the resulting image will be. We have found scale 6 to be appropriate for reducing the noise of the BCGs+ICL+background surface, yielding the image in Fig. \ref{CHEFalg}(c). Note the PMT does not conserve the enclosed total flux, which is not a problem since we have used this filtered image only for the purpose of determining the real extension of the BCG.\\
 
 We have then computed the First and Second Fundamental Forms and solved equation (\ref{curvatura}) to get the MPC map of the filtered image. The result is shown in Fig. \ref{CHEFalg}(d) and (e), as well as the curves joining the points where the MPC changed most, which we identified with the limits of the BCGs. This curve was determined by applying the k-sigma clipping algorithm to the curvature map. Segmentation of gray-scale images by thresholding the pixel intensity is a traditional technique not only used in astronomy but also in many other scientific fields (e.g. \cite{sextractor,sigmaclip1,sigmaclip2}). Based on the idea that the pixel distribution is bimodal, k-sigma clipping permits to partition the image into two segments: those pixels with larger curvature, associated with the BCG, and those with smaller curvature, related to the ICL. We find that k=1 works well for our data, as it will be proved in Section \ref{accuracy}. \\
 
\subsection{Modelling the BCGs} \label{BCGmodelling}

 Once we have marked out in the image the projected area corresponding to the BCGs, we have proceeded to fit them using the CHEFs. First, we have locally estimated the brightness level in the areas surrounding the regions defined in Sect. \ref{delimiting}, i.e., we have measure the ICL in the zones immediately enclosing the BCGs. We have then assumed this ICL as background and inserted  it into the CHEFs algorithm. In other words, we made the CHEFs ``believe'' the ICL is part of the background and therefore they fit everything down to this background level. We have thus constrained the CHEFs to fit just the BCGs without including the light from the ICL. Removing the flux corresponding to the BCGs in this way, we let the image just composed by ICL and background (see Fig. \ref{CHEFalg}(f)).\\
 
 \subsection{Estimating the ``real'' background}
 
 Although we have been able to find a geometrical rule to disentangle the light belonging to the BCGs from the ICL, it was not straightforward to find an analogous criterion to separate the background luminosity from the ICL. Moreover, the ICL is likely distributed over the whole reduced field covered by the Frontier Fields observations. Nevertheless, we have tried using the residual image in Fig. \ref{CHEFalg}(f) to look for blank areas and measure the background level from them. To identify these blank areas we have firstly filtered the noise using the B-spline Wavelet Transform, which produces a multiresolution decomposition of the image using splines as scaling functions \citep{starck}. We have chosen B-spline wavelets because they have compact support. The more compact the domain a wavelet is, the more accurate will be the decomposition of the signal. However, wavelets are subject to an analogue to Heisenberg's uncertainty principle, so a compact spatial domain is translated into a larger domain in frequency and the multiresolution information provided (i.e., the splitting into different ranges of frequency) becomes diffuse. B-spline wavelets interval of domain is not infinite, and find one of the best possible trade-offs for space-frequency localization. Moreover, splines are optimal in terms of smoothness to approximate continuous functions -which is called the minimum curvature property \citep{splines,splines2}. So, filters based on them lead to excellent denoised and smooth results.\\
 
\begin{figure}[h!]
\centering
\includegraphics[width=7.8cm]{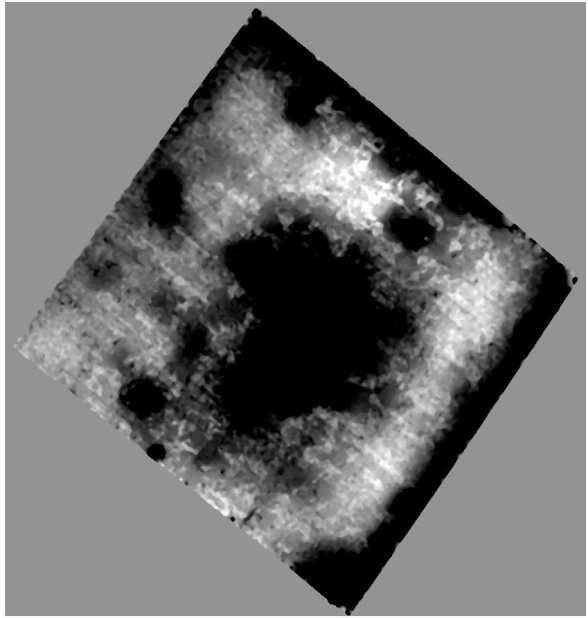}
\caption{Residual image in Fig. \ref{CHEFalg}(f) filtered with the B-spline wavelet transform to enhance the edges.}\label{residual_enhanced}
\end{figure}
 
 The B-spline wavelet filtering not only demonstrated that even the blankest areas in the image were still polluted with ICL, but was also able to make noticeable an instrumental grid-like spurious artifact (see Fig.\ref{residual_enhanced}). This cross-hatched pattern had already been noticed by \cite{artifacts} and it is believed to be caused by the drizzle process, due to correlated noise \citep{ACS_manual}. So we have adopted a different approach to estimate the background: we have searched for HST images close to Pandora, observed with the same F814W filter and observed at nearly the same epoch, to avoid time dependent systematics. We have assumed these images to have approximately the same background level as those of Abell 2744, with the advantage of being free of ICL.\\
 
\begin{figure}[t]
\centering
\includegraphics[width=7.8cm]{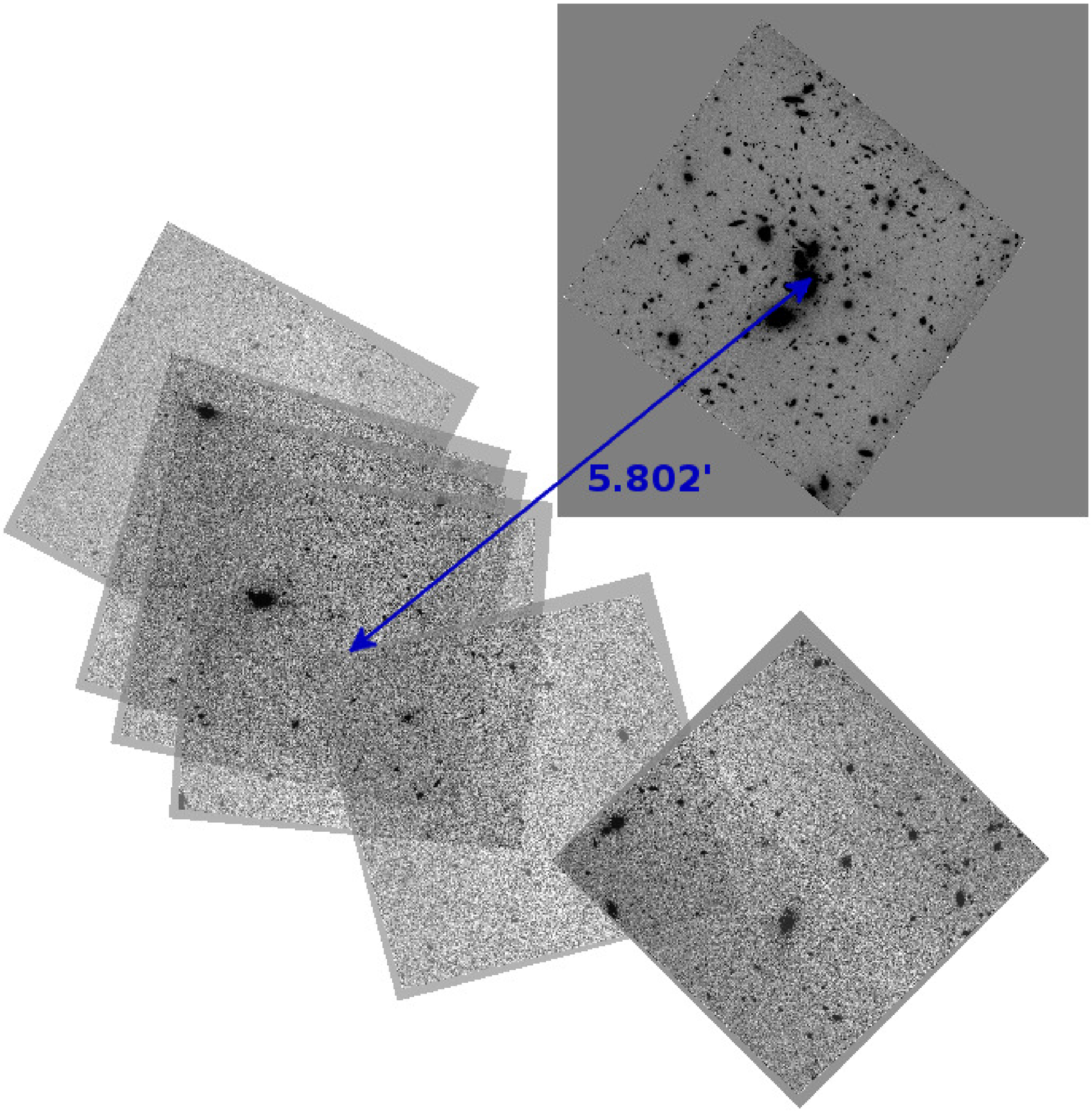}
\caption{Main cluster of the Pandora system with the seven additional pointings with similar observational characteristics found in the Hubble archive.}\label{position7images}
\end{figure}

 Using the Mikulski Archive for Space Telescopes\footnote{http://archive.stsci.edu/hst/} (MAST), we have found seven new images containing HST data of regions to the South and South-West of Abell 2744 (see Fig. \ref{position7images}). Table \ref{obs_info} compiles the dates when these images were taken and further information on their observation. The optical images of the main Pandora field were collected between 14 May 2014 and 1 July 2014, so just the first of the seven additional pointings was observed within the window of interest.\\
 
\begin{deluxetable}{cccc}
\tablewidth{0pt}
\tablehead{
\colhead{Name} & \colhead{UT date} & \colhead{Exposure time} & \colhead{\# Exposures} \\
& [mm/dd/yy] & \colhead{[seg]} & }
\startdata
jca9t3010 & 06/09/14 & 1970 & 4\\
jca9t5010 & 07/03/14 & 953 & 3\\
jca9t6010 & 07/12/14 & 923 & 3\\
jca9t7010 & 07/22/14 & 903 & 3\\
jca9t8010 & 08/17/14 & 903 & 3\\
jca503010 & 08/23/14 & 1840 & 8\\
jca504010 & 08/22/14 & 1380 & 6
\enddata
\tablecomments{Observational information of the seven additional images used to estimate the sky level.}\label{obs_info}
\end{deluxetable}

\begin{figure}
\vspace{0.2cm}
\centering
\centerline{\includegraphics[width=3.9cm]{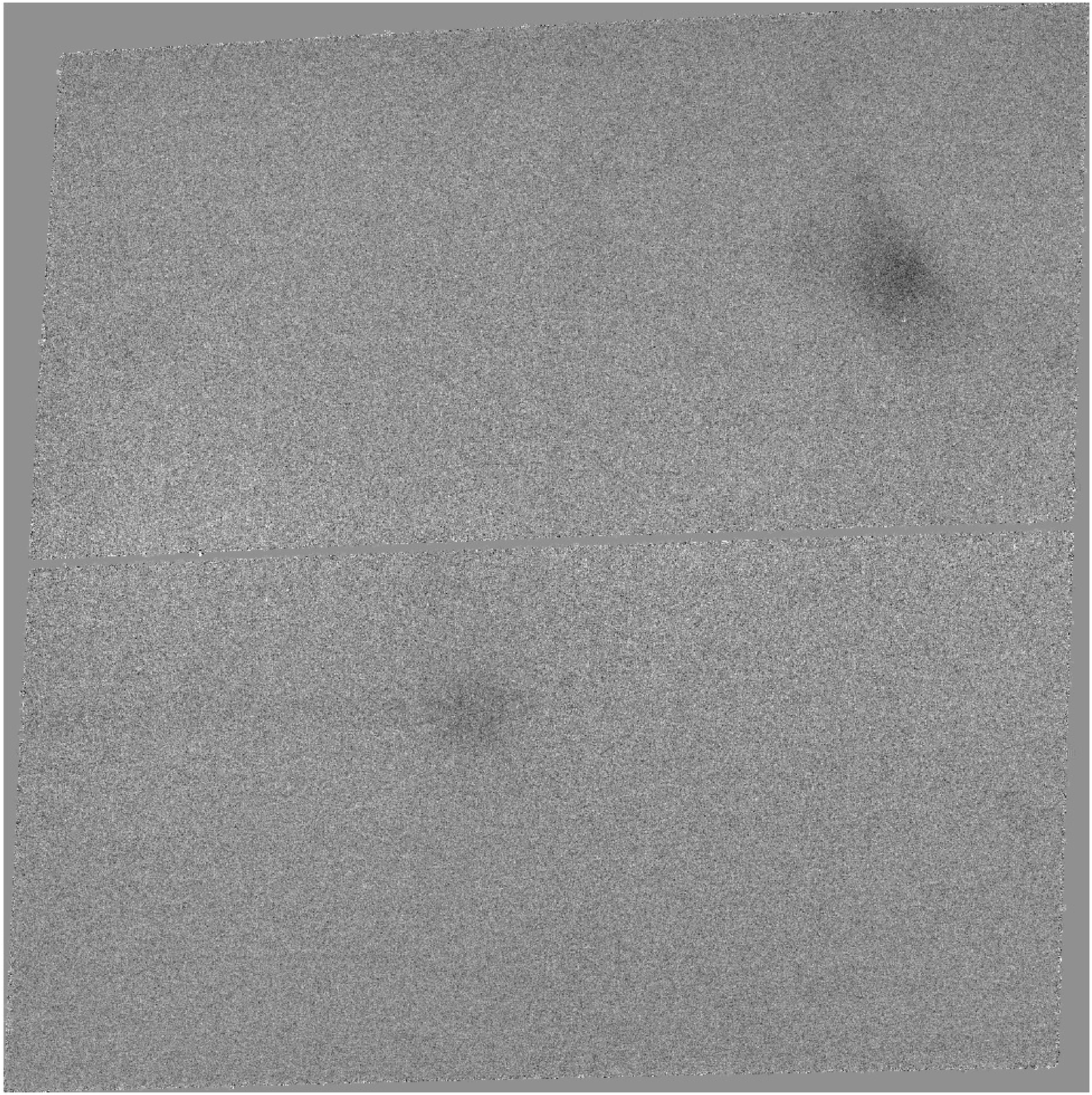}\includegraphics[width=3.9cm]{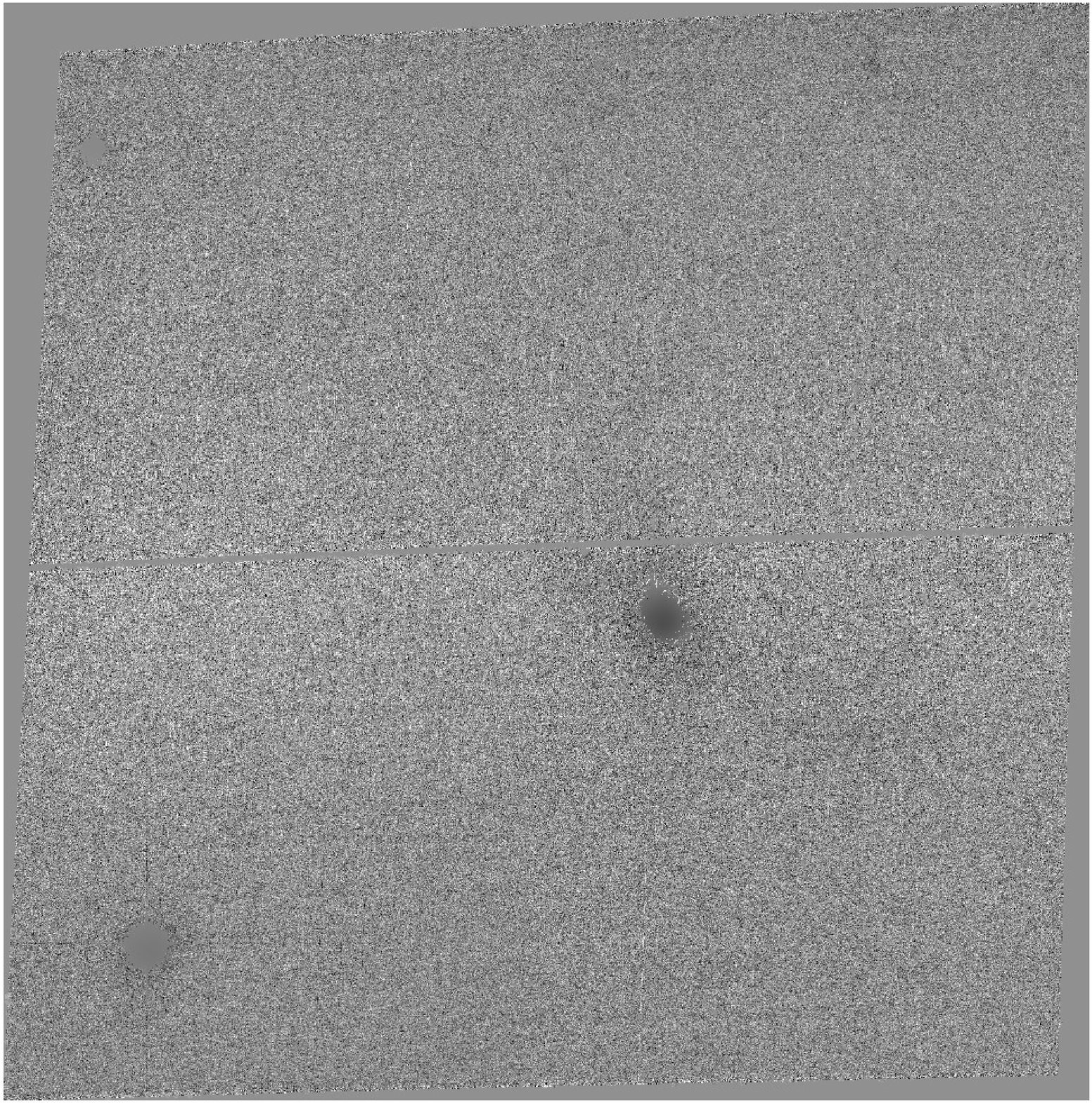}}
\centerline{\includegraphics[width=3.9cm]{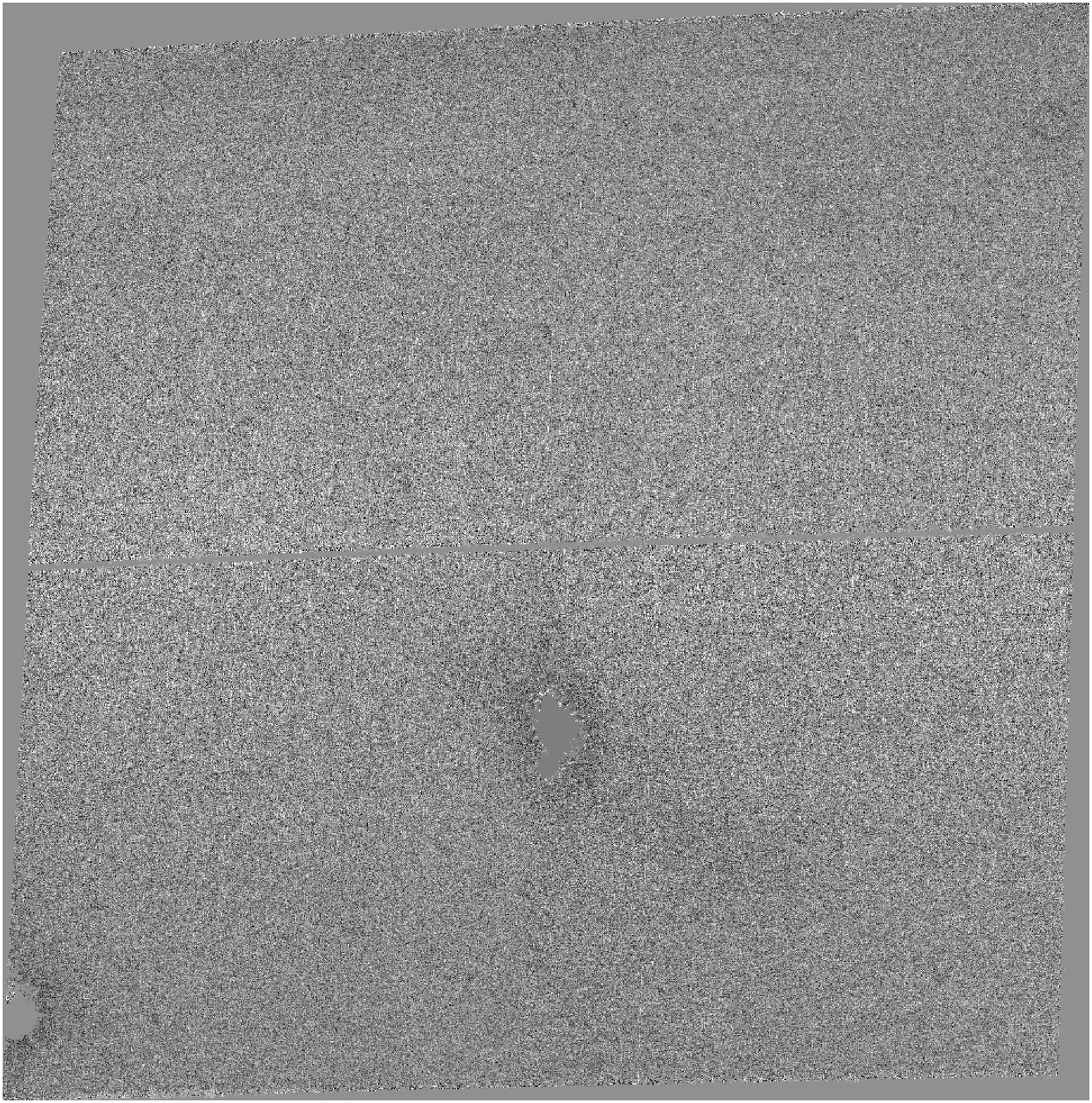}\includegraphics[width=3.9cm]{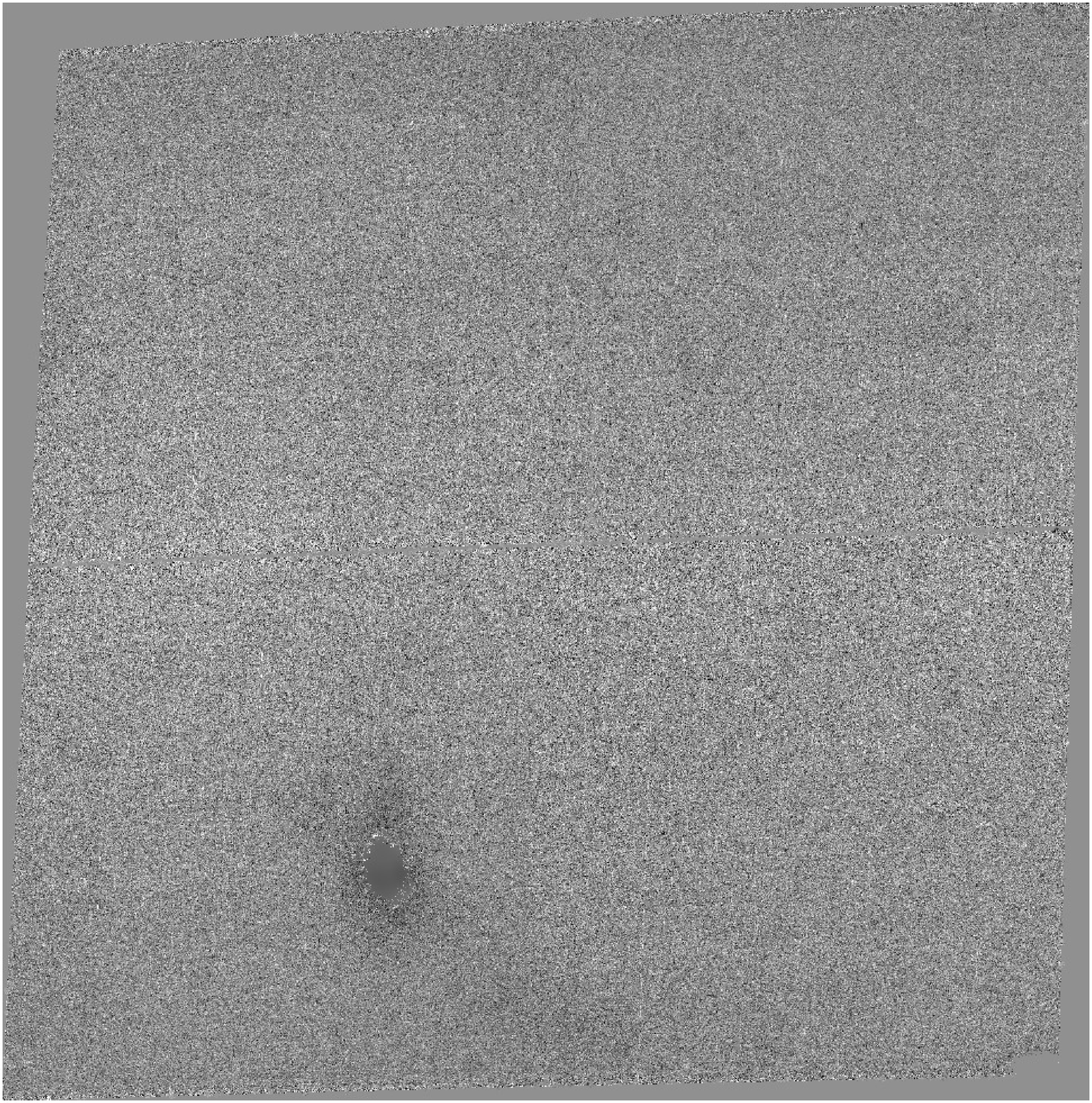}}
\centerline{\includegraphics[width=3.9cm]{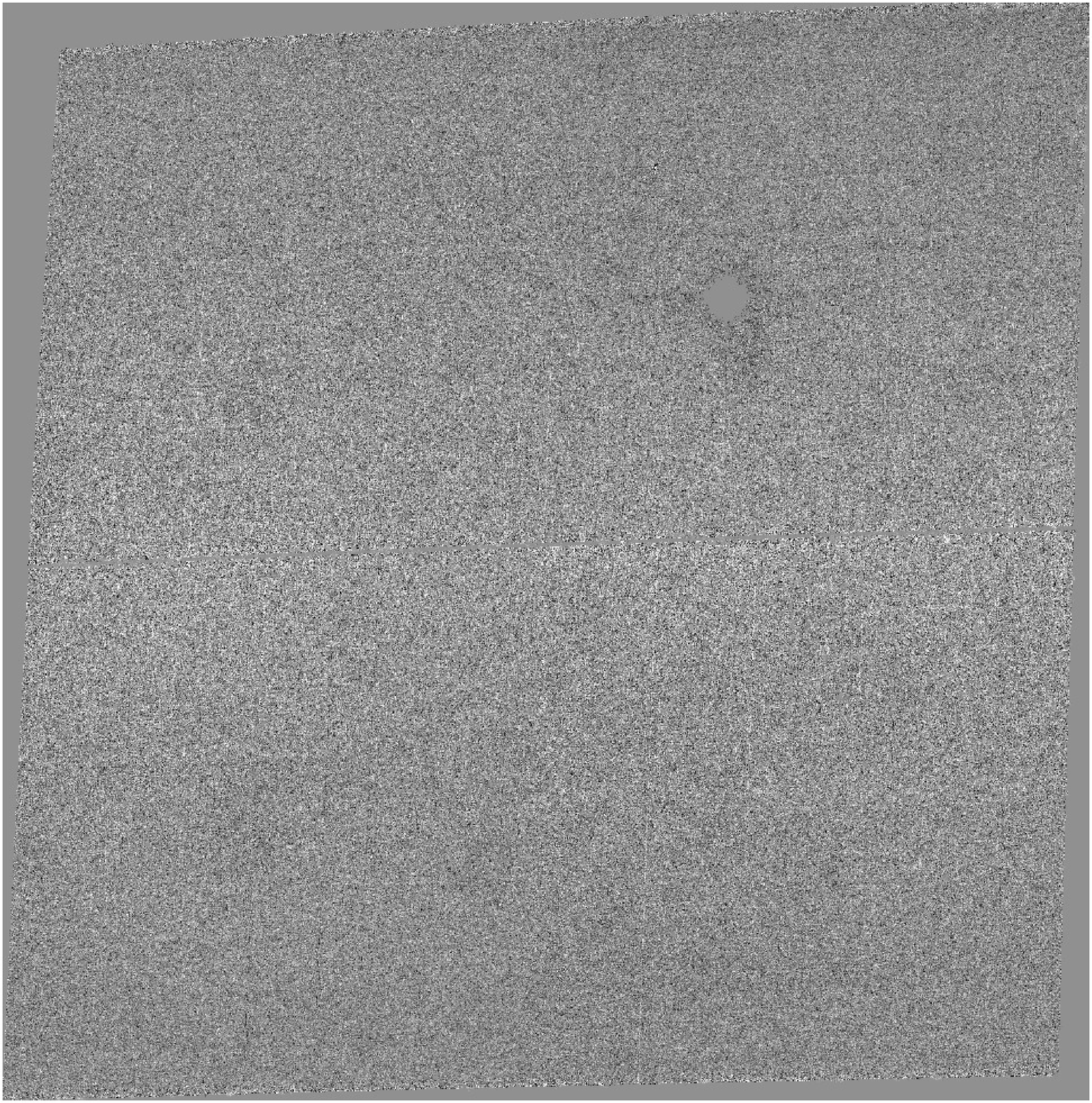}\includegraphics[width=3.9cm]{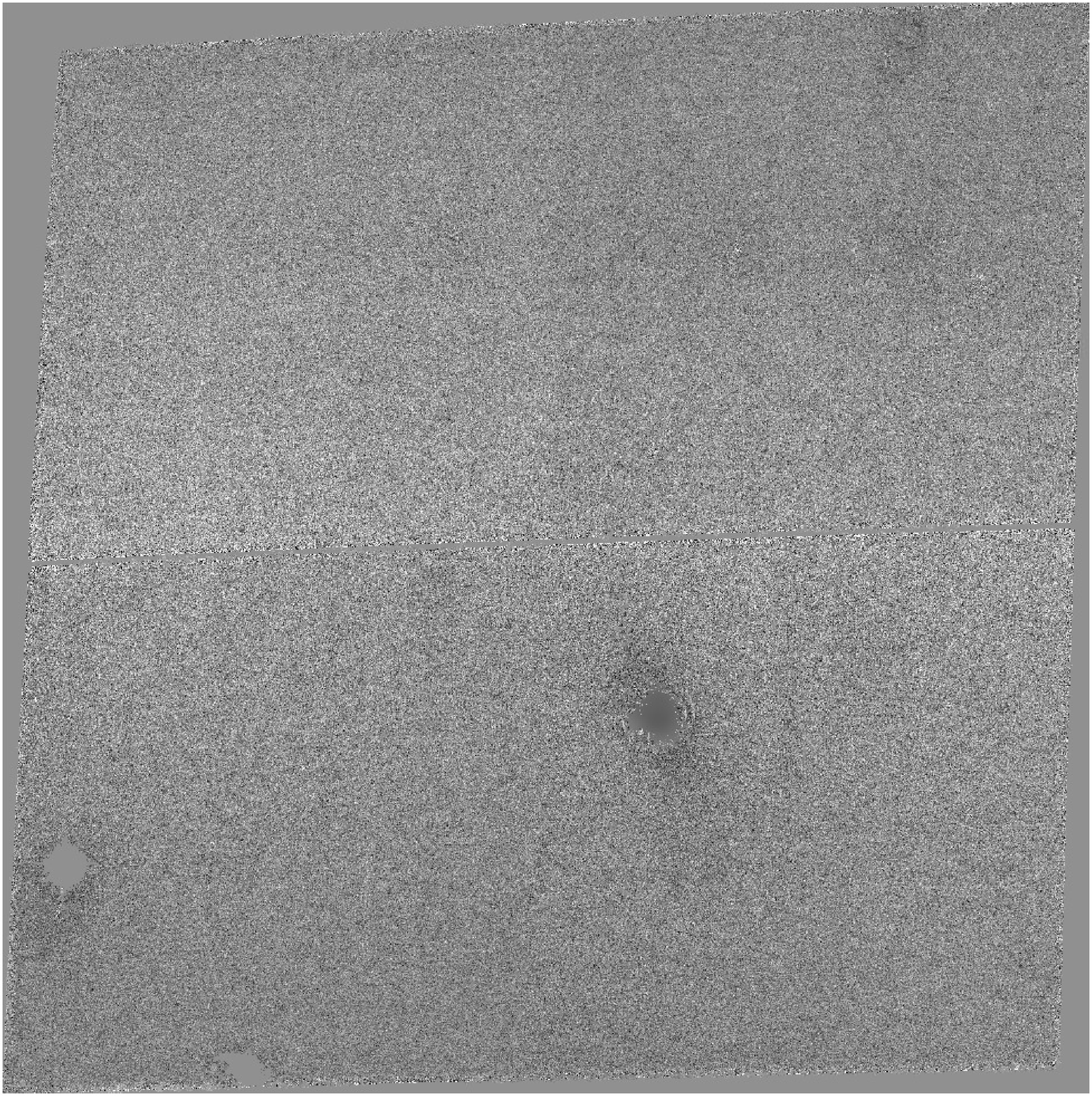}}
\centerline{\includegraphics[width=3.9cm]{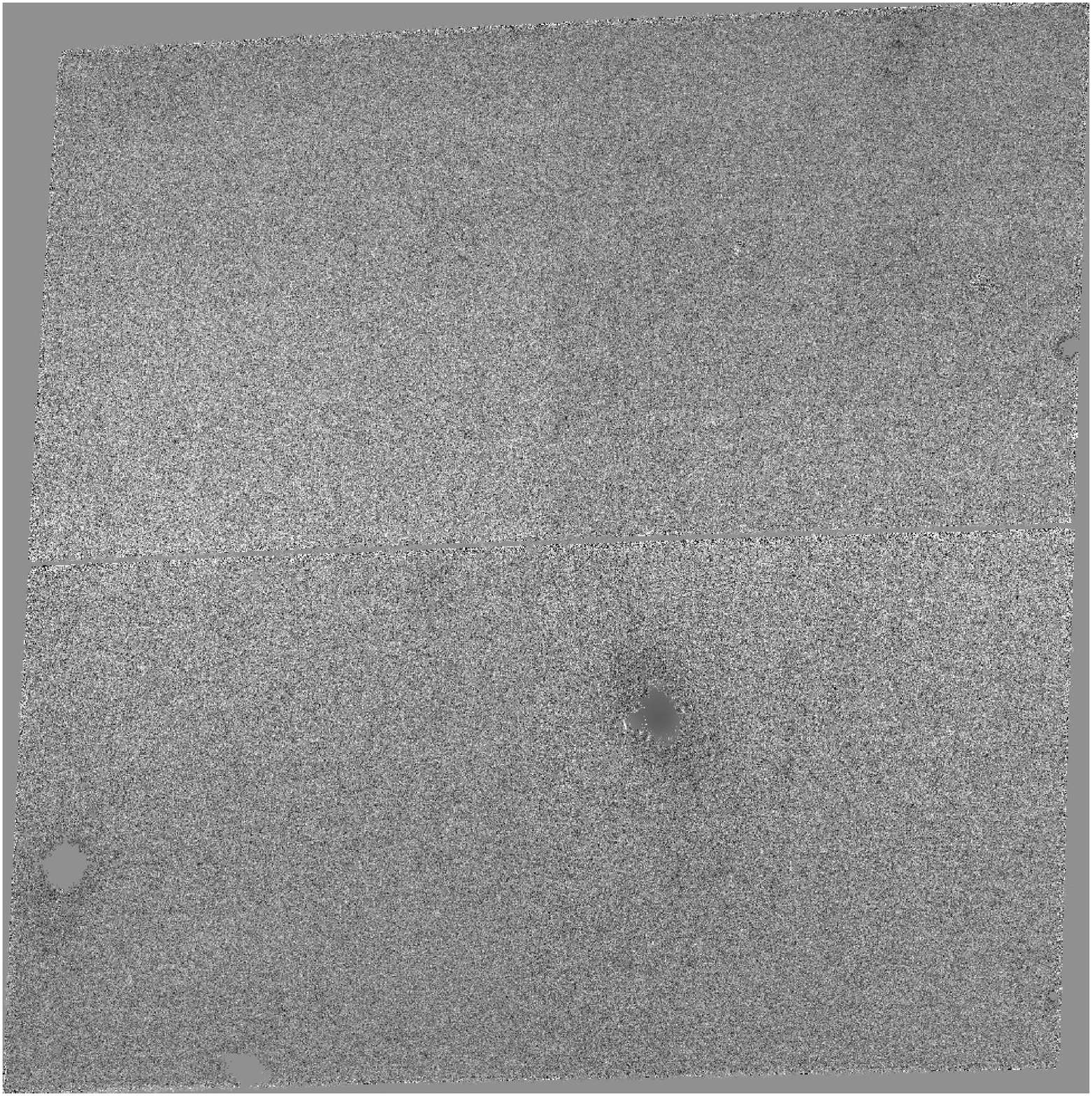}}
\caption{Residual images after subtracting the CHEF models of all the objects in the seven additional pointings to estimate the background. CHEFs failed processing two sources that were masked.}\label{residual7images}
\end{figure}

\begin{figure}
\vspace{0.2cm}
\centering
\includegraphics[width=7.5cm]{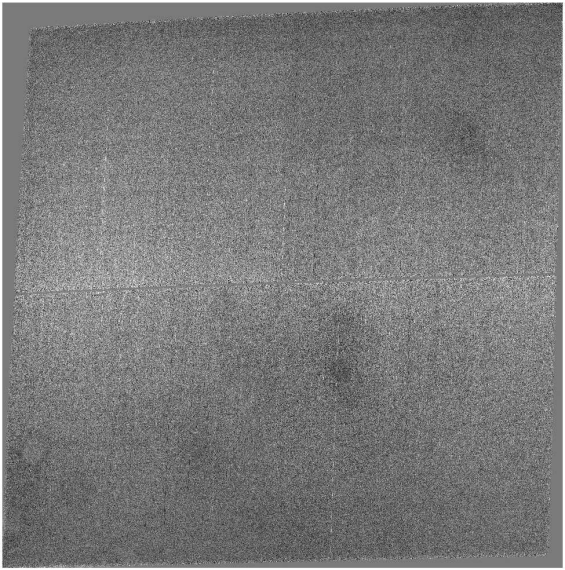}
\caption{Stack of the seven residual images in Fig. \ref{residual7images}. The cross-hatched pattern artifact due to the post-processing of the ACS data is noticeable without applying any enhance filter to increase the image contrast.}\label{stacking7images}
\end{figure}

 We have estimated the sky level individually on each of the seven images and also stacking them. We have first removed all the objects in these images using the CHEFs to later mask the sources where the CHEF fitting algorithm yielded a large chi square. The resulting clean images can be observed in Fig. \ref{residual7images} and \ref{stacking7images} (note the artifacts mentioned above become visible in this stacked image without needing to enhance the contrast). We have used these residual images to estimate a constant background level for each one of them (see Table \ref{bgs}), ignoring the error induced by the low-scale cross-hatched pattern.\\
 
\begin{deluxetable}{lcccc}
\tablewidth{0pt}
\tablehead{
\colhead{Name} & \colhead{Bg} & \colhead{Bg rms} & \colhead{ICL fraction} & \colhead{ICL fraction error}\\
& [cps] & \colhead{[cps]} & \colhead{[\%]} & \colhead{[\%]}}
\startdata
jca9t3010 & -6.1e-6 & 6e-3 & 23.27 & 4.77\\
jca9t5010 & \phantom{-}4.1e-5 & 1e-2 & 22.76 & 4.36\\
jca9t6010 & \phantom{-}4.5e-4 & 1e-2 & 17.16 & 0.47\\
jca9t7010 & \phantom{-}7.2e-5 & 1e-2 & 22.38 & 4.05\\
jca9t8010 & \phantom{-}1.3e-4 & 1e-2 & 21.60 & 3.41\\
jca503010 & \phantom{-}5.1e-4 & 9e-3 & 16.28 & 0.78\\
jca504010 & \phantom{-}4.1e-4 & 1e-2 & 17.74 & 1.07\\
Stack & \phantom{-}3.1e-4 & 4e-3 & 19.17 & 1.58\\
\enddata
\tablecomments{Background levels individually measured on the seven additional images  and corresponding ICL fractions and errors for these sky levels. A stack of the seven images is used for the final results.}\label{bgs}
\end{deluxetable}

 \subsection{Measuring the ICL} \label{measureICL}
 
\begin{figure}
\vspace{0.2cm}
\centering
\includegraphics[width=7.5cm]{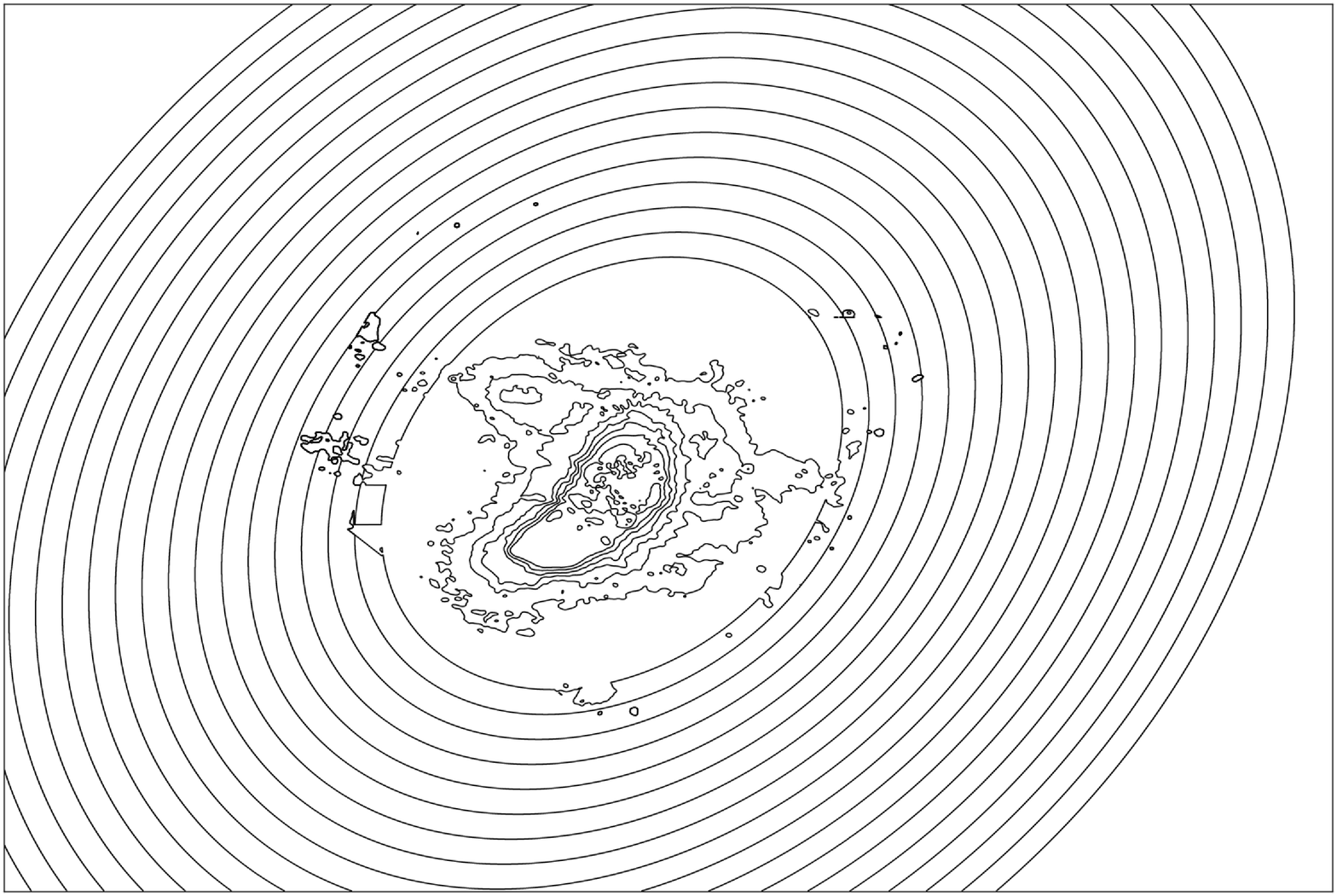}
\caption{ICL contour regions to measure the profiles in Fig. \ref{profiles}. Contours in the central area fit the ICL silhouette whereas ellipses are chosen for the outer zones.}\label{contours}
\end{figure}

 Once we have estimated a background level, we measured the ICL fraction splitting the image in regions delimited by the natural contours of the ICL (inner area) and ellipses (outer area), as shown in Fig. \ref{contours}. As it can be observed in Fig. \ref{residual_enhanced}, a non-negligible amount of light remained in the borders of the image after removing all the objects. This flux is clearly unrelated to the ICL of the main cluster of Pandora, since it is caused by the image reduction process. We could not use the natural contours throughout the residual image to measure the ICL since they were delimiting this extra light in the outer areas of the image. Ellipses were instead used in the periphery, since they seemed to be the best regular geometrical shape enclosing the natural contours.  The axes ratio and position angle of the ellipses were chosen to best fit the largest inner contour line, and the semimajor axes were equally spaced.\\
 
\begin{figure}[h!]
\centering
\vspace{0.2cm}
\includegraphics[width=7.8cm]{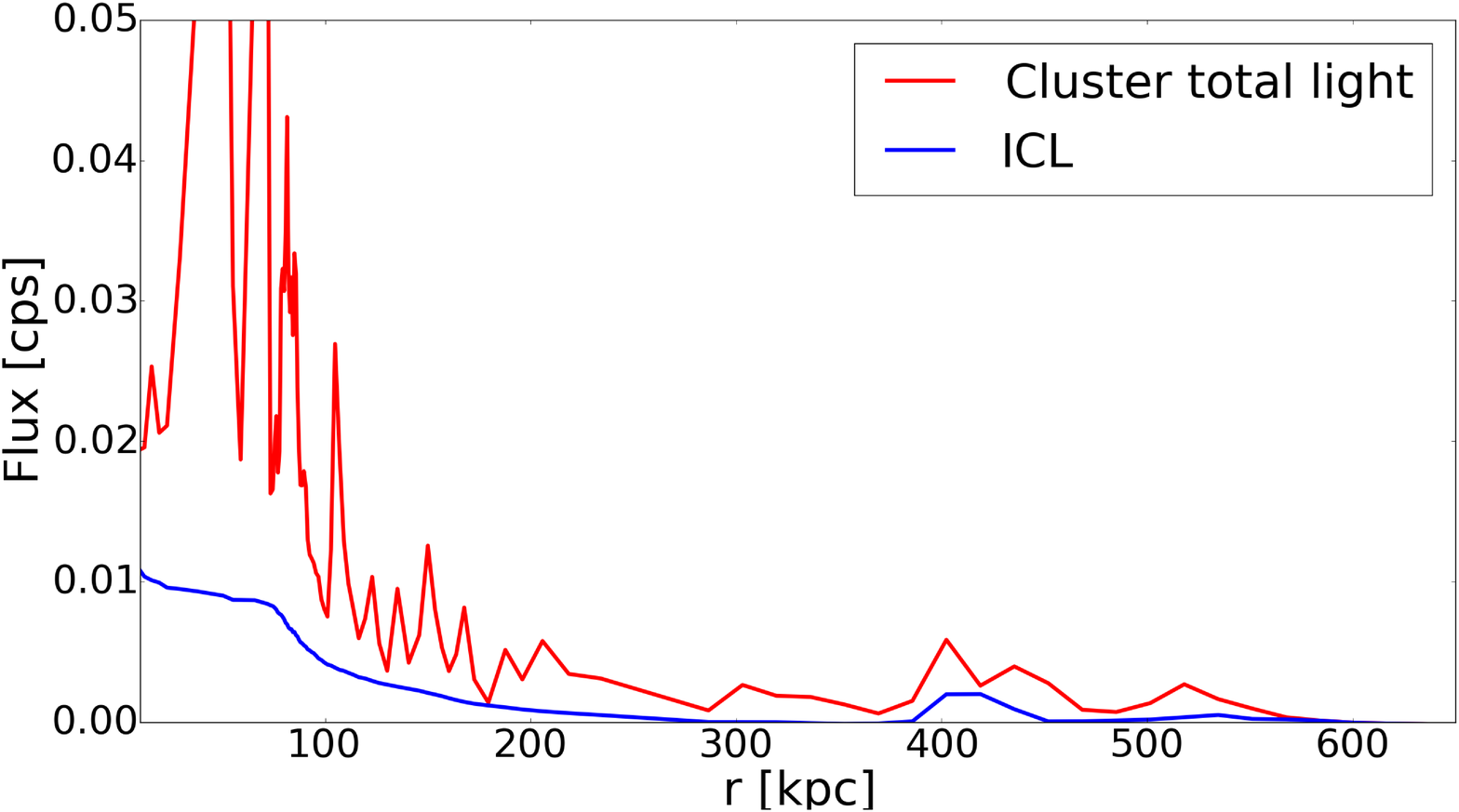}
\caption{Radial profiles of the ICL (blue) and the total cluster light (red), using the contours in Fig. \ref{contours} divided into eight angular regions. The apex of total cluster profile is not reached at radius 0 since the contours fit the ICL, so the two peaks at ~41 and ~84 kpc correspond to the two brightest galaxies in the cluster.}\label{profiles}
\end{figure}
 
 We have calculated the radial profiles for both the ICL and the total cluster light measuring the flux within these apertures (see Fig. \ref{profiles}). We have defined $r$ in that plot as the radius of the circular aperture such that its area is equivalent to the region enclosed by each contour. The minimum in the ICL radial profile defines the point where the ICL submerges in background, thus, the region to measure the ICL contribution to the total cluster light. We found an ICL equivalent radius of ~353 kpc and a final ICL fraction of 19.17$\pm$1.58\% within this area, using the background level estimated from the stack of the seven additional images. In Table \ref{bgs} we also include the ICL fractions computed with the sky levels of the individual images for comparison. The resulting values range from ~16\% to ~23\% depending on the background level estimated. The ICL fraction errors are estimated as described in \cite{8regions}, \cite{zibetti}, and \cite{presotto}, dividing each contour region into eight angular ``pie'' sections and measuring the ICL and its rms from each sector. Differently from these works, instead of splitting the contour regions in eight equal sections, we have made them have the same area to obtain uniform statistics.\\
 
\section{Accuracy of the method} \label{accuracy}

 The 1.58\% error obtained in last section is the error associated to the precision of the data, calculated empirically by splitting the contour regions into pie sections. However, the method described in this work also has inherent uncertainties that must be incorporated. To calculate the accuracy of our algorithm, we generated several artificial fields containing a circular BCG and ICL with simple analytical profiles. The simulated images display a wide variety of configurations for the BCG+ICL system, with different steepness and noise levels. We conducted two tests: the first to estimate the accuracy of the method as a function of the smoothness of the BCG-to-ICL transition, and the second to analyse the effect of noise in the final ICL fraction.\\
 
\subsection{BCG-to-ICL transition test}

 We wanted to know the sensitivity of our method to the shape of the BCG+ICL surface. The ability of the MPC parameter to distinguish the different steepness of the BCG and the ICL light distributions determines the accuracy of our algorithm. We would expect the accuracy to decrease as the profiles of the two components resemble each other and the transition from the BCG to the ICL is smoother. To analyse and quantify this, we generated a set of mock images using double exponential profiles as in the simulations by \cite{darocha}. This simple configuration has also been described in many observational studies (e.g. \cite{krick}).\\
 
\begin{figure*}[t!]
\centering
\includegraphics[width=16.5cm]{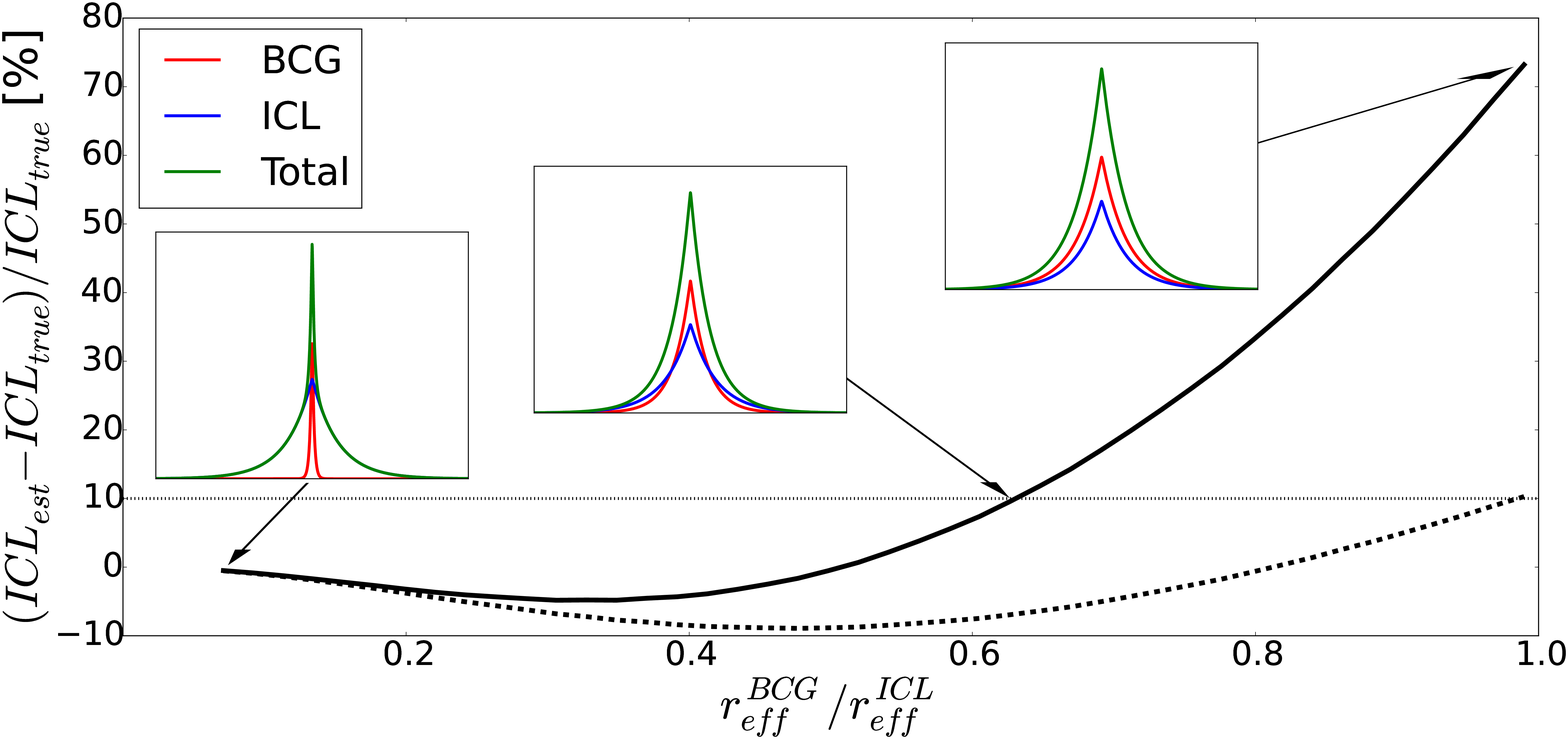}
\caption{ Results of the test with several BCG-to-ICL transitions. The relative offset between the estimated and real ICL fluxes is represented, versus the ratio of the BCG and ICL effective radii (black solid line). The corrected relative offset, after removing the effect of the BCG outermost flux is represented with a black dashed line. We also show the radial profiles of the BCG (red), the ICL (blue), and the total surface (green) for three of the simulated fields, corresponding to the limiting cases of a very steep BCG (left inset) and an ICL completely embedded into the BCG flux (right inset), as well as the intermediate case which corresponds to a 10\% error (middle inset).}\label{test1}
\end{figure*}

 The exponential profile has two free parameters: the effective radius and the central surface brightness \citep{peng}. For the first of our simulated images we set these parameters as described in \cite{zibetti} for an average cluster in the redshift range z$\sim$0.2-0.3. The BCG had thus an effective radius of 19.29 kpc ($\sim$5 arcsec at z$\sim$0.25) and the ICL's effective radius was 275 kpc ($\sim$70 arcsec at z$\sim$0.25). Despite the fact that Abell 2744 data used in this work were observed by the HST, we chose to simulate the fields with SDSS resolution for the sake of computational speed. As this ICL effective radius covers 180 pixels with SDSS pixel scale, we evaluated the exponential profiles on a grid of 1500x1500 pixels$^2$ to ensure all the flux from the ICL was included. As also described in \cite{zibetti}, we adopted a surface brightness of 27.5 mag/arcsec$^2$ at 100 kpc radius for the ICL and 24 mag/arcsec$^2$ for the BCG. This artificial image thus consists on a very steep BCG and a extended ICL (left inner frame in Fig. \ref{test1} shows the radial profile of this system, with the BCG plotted in red, the ICL in blue, and the final composite surface in green).\\
 
 We used this first image as a prototype to generate other mock fields. We kept the ICL surface distribution fixed and varied the BCG profile increasing its effective radius gradually. Middle and right insets in Fig. \ref{test1} show examples of the new simulated images: an intermediate case where the BCG and ICL slopes are very similar to one another, and the extreme case where their effective radii coincide, respectively. Our final set contained 44 simulated images, characterized by the ratio of the BCG effective radius to that of the ICL, $r_{eff}^{BCG}/r_{eff}^{ICL}$. This ratio, called BCG-to-ICL ratio from now on, spans the interval [0.07,1] in our simulations. As the BCG-to-ICL ratio approached unity, the more indistinguishable the transition between the BCG and ICL surface distributions became (see insets in Fig. \ref{test1}). Applying our method to these simulations allowed us to know the upper limit in the BCG-to-ICL transition to which this method can be applied, as well as its intrinsic scatter. Note that for other techniques of ICL measurement this applicability limit also exists although is not usually stated (e.g. fitting with two traditional profiles or multiscale decomposition with wavelets).\\
  
 We then applied the CHEF technique to each one of the 44 simulated fields, as we did for Abell 2744. Just the denoising using the PMT was skipped, since simulations were purposely free of noise (we will explore the effect of the noise in the next section). To quantify the accuracy of our technique, we calculated the difference between the real ICL flux and that estimated by our algorithm, normalized by the real ICL flux. Hereafter, this quantity is called relative offset. Fig. \ref{test1} summarizes the results, plotted with a black solid line. The relative offset is almost null for the first mock image, the one described as the average cluster for z$\sim$0.2-0.3 in \cite{zibetti}. We then observe two different behaviours in the relative offset: up to 
$r_{eff}^{BCG}/r_{eff}^{ICL}\sim 0.5$ it is negative with a minimum relative offset of $\sim$-4.8\%, and positive otherwise. That means that the algorithm slightly underestimates the ICL flux in the images with steeper BCG profiles, showing a downward trend up to $r_{eff}^{BCG}/r_{eff}^{ICL}\sim 0.35$. From that BCG-to-ICL ratio there is an upward trend that makes the relative offset become positive from $r_{eff}^{BCG}/r_{eff}^{ICL}\sim 0.5$ on. Thus, when the effective radii of the BCG and the ICL are more similar, our method overestimates the ICL flux. This error increases as the BCG-to-ICL transition is smoother, as expected. For the extreme cases where the BCG-to-ICL ratio is close to 1, the BCG surface distribution completely includes the ICL profile, it is not possible to disentangle them and the offset is very high. We mark with arrows the relative offsets that correspond to the mock radial profiles displayed as examples, and we added a black dotted line to point out the 10\% relative offset limit. For the mock field that corresponds to this offset (middle inset) the profiles of the BCG and the ICL are already almost indistinguishable.\\
 
 Throughout this paper we explained how the curvature parameter marks out what we called the ``real'' extension of the BCG. However, the MPC delimits the BCG-dominated area of the total light distribution. The BCG actually extends beyond this region, although the ICL contribution to the total profile is larger. Since the CHEFs are constrained to the BCG-dominated area, as described in Section \ref{BCGmodelling}, the BCG is modelled down to a background level outside this region. The light from the outermost regions of the BCG is neglected, thus underestimating the BCG flux and overestimating the ICL. For the first simulated fields, this effect is negligible, since the BCG is very steep and the flux contained in the wings of the BCG is very small. The relative offset associated to these mock images, in practice, represents the accuracy of the method for detecting the transition from the BCG to the ICL. From ratio $r_{eff}^{BCG}/r_{eff}^{ICL}\sim 0.35$, the upwards trend in the relative offset shows that the effect of the BCG peripheral light becomes significant. In the interval [0.35,0.5], the relative offset is still negative, which indicates that there is a trade-off between the overestimation of the ICL caused by the BCG wings, and the underestimation derived from the inaccuracy in the BCG demarcation. From $r_{eff}^{BCG}/r_{eff}^{ICL}\sim 0.5$, the relative offset becomes positive, so it is dominated by the flux in the BCG wings. Trivially, the offset increases as the wings are more extended.\\
 
 We ran a second test to separate the effects of the two sources of error (although they are not completely independent; the amount of BCG flux that is neglected depends on how accurately the BCG-to-ICL transition is determined). For each simulated field we measured the flux contained in part of the BCG the wings that lay on the ICL-dominated region. We subtracted it from the ICL flux estimated by our algorithm. Although this cannot be done with real data, running this simple test on mocks allowed us to estimate the accuracy of our algorithm to disentangle the light components, almost independently of the precision of the BCG model that is later calculated. The dashed back line in Fig. \ref{test1} represents the new relative offset after the BCG wings correction described above is performed. The effect of the outermost BCG light is still visible, although much smoother. The trend change occurs at ratio $r_{eff}^{BCG}/r_{eff}^{ICL}\sim0.48$ and the consequent overestimation is much lower. For every configuration of the BCG+ICL system, the error is less than 10\% (in module). Notice that other techniques of ICL measurement also neglect the flux from the BCG wings, such as the surface brightness thresholding or wavelets.\\
 
 For the particular case of Abell 2744, we used the software GALFIT \citep{peng,peng2} to fit traditional profiles and estimate the effective radius of the two BCGs and the ICL. Specifically, we used three exponential profiles to model the composite surface, plus a constant component for the background. All the profiles were fitted simultaneously, yielding the following effective radii: $r_{eff}^{BCGs}=20.0$ and $15.2$ pixels for the BCGs and $r_{eff}^{ICL}=130.6$ pixels for the ICL. The corresponding BCG-to-ICL ratios are thus $r_{eff}^{BCGs}/r_{eff}^{ICL}=0.15$ and $0.12$, which correspond to offsets lower than 2\% (in module), according to Fig. \ref{test1}.\\
 
\subsection{Noise test}

 We wanted to estimate the effect of noise in the accuracy of our method, as a function of the S/N. The idea of this test is selecting one of the simulated fields described in the previous section and polluting it with Poissonian sky noise. To mimic the characteristics of our real data from Pandora we ran this test twice, using the mock images with BCG-to-ICL ratios equal to the BCGs observed in Pandora. We thus used the simulations with ratio $r_{eff}^{BCG}/r_{eff}^{ICL}=0.15$ and $0.12$ as prototype images to generate the two sets of noisy mock fields. To get different levels of S/N, we varied the surface brightness of the ICL in the prototypes before adding the Poisson noise. The surface brightness of both simulation sets spanned the interval [24.5,37] mag/arcsec$^2$. The corresponding S/N after including the noise obviously depended on the flux of the BCG, but it overall ranged from 0.1 to 35 approximately, for both sets of simulations. We applied exactly the same technique that was described for Abell 2744, including the denoising of the images with the PMT.\\
 
 \begin{figure}[h]
\centering
\includegraphics[width=7.8cm]{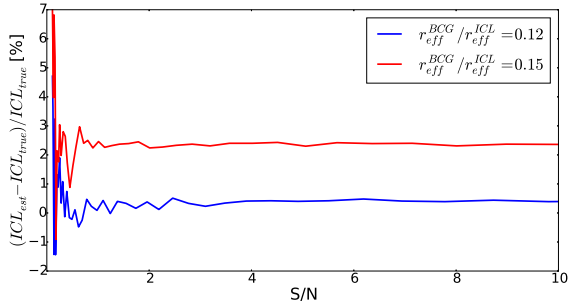}
\caption{Results of the noise test. The relative offset between the estimated and real ICL fluxes is represented for each simulated field, versus the S/N. Two different configurations for the BCG+ICL system were chosen, resembling the two BCGs in Pandora cluster, with $r_{eff}^{BCG}/r_{eff}^{ICL}=0.12$ (blue line) and 0.15 (red line).}\label{test2}
\end{figure}
 
 Although we ran this test up to $S/N\sim 35$, we show the results up to $S/N\sim 10$ for the sake of clarity since there is virtually no change after that (Fig. \ref{test2}). We plot the obtained relative offsets with different colours for each set of simulations. As expected, the accuracy decays for very low values of S/N. For the two sets, the offset is significant up to $S/N\sim 0.2$, with a maximum 7\% offset. From this point, the accuracy of the method improves to finish converging to fixed relative offsets of 0.41\% and 2.37\%  for $r_{eff}^{BCG}/r_{eff}^{ICL}=0.12$ and $0.15$, respectively. Thus, the transition curve from the BCG- to the ICL-dominated region is indeed very well determined even in the presence of noise, and both the curvature parameter and the denoising algorithm work very well. We estimated that the BCGs+ICL system in the Pandora images observed by the Frontier Fields program has $S/N\sim 30.4$, using the residual image in Fig. \ref{CHEFalg}(b) and its denoised version (Fig. \ref{CHEFalg}(c)). For this S/N we obtained relative offsets of 0.42\% and 2.37\%, respectively for each BCG. Adding these offsets in quadrature to the 1.58\% error empirically obtained in Sect. \ref{measureICL}, we estimated a final ICL fraction error of 2.87\%. \\
 
\section{Conclusions}\label{conclusions}

 Given the myriad of methods to measure the ICL fraction existent in the literature and the disparity in the results \citep{rudick}, we decided to develop a completely independent new technique using the CHEFs, which is more robust and less dependent on arbitrary choices about galaxy's light distribution and background level. Other previous techniques in the literature are either not precise enough or rely on different assumptions on the morphology, the surface brightness, or the density of the ICL that can introduce systematics and bias in the estimation of the ICL fraction. The only hypothesis we assume in our method is that the intensity curvature profile of the BCGs must be different from that of the ICL, so we can disentangle the limits of the BCG brightness from the ICL. We have tested this algorithm using the recently published data from Abell 2744, observed by the Frontier Fields program. These data represent the deepest images ever collected of this merging, highly complex cluster.\\
 
 We have used the CHEFs to subtract the galaxy light contribution at several stages of the process; removing firstly all the foreground and background galaxies and later the brightest galaxies around the center of the cluster. To precisely mark out the extension of these BCGs, we have used a surface curvature parameter from differential geometry, previously smoothing the image with a multiscale transform.\\
 
 Given that the FoV of the Frontier Field images of Pandora is relatively small and likely to be fully contaminated with cluster light, separating the ICL from background directly was not possible. Therefore, to bypass this difficulty, we have looked for new images close to the main cluster of the Pandora system but free of ICL, and with similar observational characteristics. We have found seven new images in the Hubble archive, nearly 6 arcmin far from the center of the cluster, observed with filter F814W approximately at the same epoch. We have estimated the background level not only individually but also from a stack of these additional images. \\
 
 We have completed the redshift catalog from \cite{owers} with the available information in NED to determine the cluster membership. Interlopers have been identified and rejected with the PEAK and the shifting gapper methods, in a two-step analyses. The estimated ICL fraction is 19.17$\pm$2.87\%, which is very consistent with numerical simulations that predict an ICL fraction between 6\%-24\% for a cluster at redshift z$\sim$0.3 \citep{trujillo}. It is not straightforward to compare this result with other ICL estimations found in the literature due to the disparity of methods and different hypotheses assumed. \cite{krick07} calculated ICL fractions of 14$\pm$5\% in $r$ band and 11$\pm$5\% in $B$ using ground-based images, which is in good agreement with our results. \cite{zibetti} analyzed one of the largest samples of clusters ever studied, with a total of 683 galaxy clusters between z$\sim$0.2-0.3, being our result very consistent with this work (see Fig. 8  in \cite{zibetti}). However, our ICL fraction is considerably higher than the ones estimated by \cite{trujillo} (considering a constant M/L ratio), using exactly the same images from the Frontier Fields program. These fractions range from 4\% to 10.5\% depending on the parameter used to derive them (stellar mass density, surface brightness, or radial distance). Although the authors claimed these values were lower limits, such a low ICL fraction would not be expected in a dynamically very active, merging system as Pandora, at redshift z$\sim$0.3064. Several studies correlated the infall activity on clusters with the quantity of optical ICL, which would be boosted by the shredding of the interacting, merging structures \cite[e.g.][]{darocha,pierini,adami}. The multiplicity of BCGs \citep{krick07}, the assymetry of the ICL, the mismatch between the ICL and the X-ray morphology \citep{pierini}, the substructures in the X-ray emission \citep{owers,merten}, the bluer color of the ICL than the BCGs \citep{pierini, trujillo}, the non-Gaussianity of the velocity distribution and the several substructures present in it \citep{boschin}, and the presence of tidal streams \citep{pierini} in A2744 support the hypothesis that the system is undergoing strong merging events. Thus, an ICL fraction of 19.17$\pm$2.87\% is another fact that supports the active dynamical state of the Pandora system, and also backs our new technique of measuring the ICL in comparison with the traditional ones.\\

\end{document}